\documentclass[iop,numberedappendix,twocolappendix]{emulateapj}

\usepackage{natbib}
\usepackage{subfigure}
\usepackage{graphicx}

\def\e{\mathrm{e}}

\newcommand{\bracket}[1]{\left<#1\right>}

\begin{document}

\title{The criterion of supernova explosion revisited: the mass
  accretion history}

\author{
Yudai Suwa\altaffilmark{1,2},
Shoichi Yamada\altaffilmark{3,4},
Tomoya Takiwaki\altaffilmark{5},
and 
Kei Kotake\altaffilmark{6,5}
}

\altaffiltext{1}{Yukawa Institute for Theoretical Physics, Kyoto
  University, Oiwake-cho, Kitashirakawa, Sakyo-ku, Kyoto, 606-8502,
  Japan}

\altaffiltext{2}{Max-Planck-Institut f\"ur Astrophysik,
  Karl-Schwarzschild-Str. 1, D-85748 Garching, Germany}

\altaffiltext{3}{Department of Physics, Waseda University, 3-4-1
  Okubo, Shinjuku, Tokyo 169-8555, Japan}

\altaffiltext{4}{Advanced Research Institute for Science \&
  Engineering, Waseda University, 3-4-1 Okubo, Shinjuku, Tokyo
  169-8555, Japan}

\altaffiltext{5}{Center for Computational Astrophysics, National
  Astronomical Observatory of Japan, Mitaka, Tokyo 181-8588, Japan}

\altaffiltext{6}{Department of Applied Physics, Fukuoka University,
  Fukuoka 814-0180, Japan}

\begin{abstract}
By performing neutrino-radiation hydrodynamic simulations in spherical
symmetry (1D) and axial symmetry (2D) with different progenitor models
by \cite{woos07} from 12 $M_{\odot}$ to 100 $M_{\odot}$, we find that
all 1D runs fail to produce an explosion and several 2D runs
succeed. The difference in the shock evolutions for different
progenitors can be interpreted by the difference in their mass
accretion histories, which are in turn determined by the density
structures of progenitors. The mass accretion history has two phases
in the majority of the models: the earlier phase in which the mass
accretion rate is high and rapidly decreasing and the later phase with
a low and almost constant accretion rate. They are separated by the
so-called turning point, the origin of which is a change of the
accreting layer. We argue that shock revival will most likely occur
around the turning point and hence that its location in the $\dot
M$-$L_\nu$ plane will be a good measure for the possibility of shock
revival: if the turning point lies above the critical curve and the
system stays there for a long time, shock revival will obtain. In
addition, we develop a phenomenological model to approximately
evaluate the trajectories in the $\dot M$-$L_\nu$ plane, which, after
calibrating free parameters by a small number of 1D simulations,
reproduces the location of the turning point reasonably well by using
the initial density structure of progenitor alone. We suggest the
application of the phenomenological model to a large collection of
progenitors in order to infer without simulations which ones are more
likely to explode.
\end{abstract}


\section{Introduction}
\label{sec:intro}

Core-collapse supernova is one of the most energetic explosions in the
universe. Although the explosion mechanism are yet to be uncovered,
there are a few possible models. Among them, neutrino heating
mechanism \citep{beth85} is the most promising scenario, in which
copious neutrinos are emitted in the vicinity of protoneutron star
(PNS) and are partially absorbed by postshock material. In this
system, neutrinos transfer internal energy from inside to the outside
of PNS and act effectively as a heating source for the postshock
layer.

Although this neutrino-heating mechanism certainly works,
state-of-the-art simulations of neutrino-radiation hydrodynamics
cannot produce an explosion in spherical symmetry
\citep{ramp00,lieb01,thom03,sumi05}. Recently, modern
multi-dimensional simulations became possible and several exploding
simulations have been reported (e.g.,
\citealt{bura06a,mare09,suwa10,muel12b,brue13,pan15} in two dimensions
(2D) and \citealt{taki12,mels15a,lent15,muel15b} in three dimensions
(3D)). These simulations were limited to a relatively small number of
progenitors:\footnote{See also \citet{kota12}, which includes the
  spherically symmetric simulations.}
\begin{itemize}
\item 11.2 $M_\odot$ of \citet{woos02}:
  \citet{bura06a,mare09,taki12,muel12b,suwa13b,suwa14a,taki14,muel15b}
\item 13 $M_\odot$ of \cite{nomo88}: \cite{suwa10,suwa11b}
\item 15 $M_\odot$ of \cite{woos95}:
  \citet{bura06a,mare09,suwa11b,muel12b,suwa13b}
\item 25 $M_\odot$ of \cite{woos02}: \citet{suwa11b,muel13}
\item 27 $M_\odot$ of \citet{woos02}: \citet{muel12c,pan15}
\end{itemize}
Other progenitor models, 8.1 $M_\odot$ \citep{muel12c} and 9.6
$M_\odot$ \citep{muel13,mels15a}, were also investigated.  More
recently, \cite{brue13} performed a systematic study using a
progenitor series of \cite{woos07} from 12 $M_{\odot}$ to 25
$M_{\odot}$ and found similar explosions for all progenitors.
\cite{dole15} reported, however, that they found that none of them
resulted in an explosion.\footnote{In addition, \citet{burr06} and
  \citet{ott08} also found that the neutrino heating is not enough to
  produce explosion.}
More recently, \cite{mels15b} performed 2D and 3D simulations with 20
$M_{\odot}$ of the same series and found an explosion in 2D and
failure in 3D with standard neutrino opacities. In this study, we
perform two-dimensional simulations for a broader mass range from 12
$M_\odot$ to 100 $M_{\odot}$ using the same progenitor series of
\citet{woos07}.

The progenitor structure is one of the most important ingredients in
the core-collapse supernova explosion mechanism\footnote{There are, of
  course, many other important issues raised so far. The most
  high-profile for the moment is dimensionality of hydrodynamics
  \citep[e.g.][]{ohni06,murp08,nord10,hank12,burr12,couc13b,taki14}.
  See also \cite{suwa13b} and \cite{couc13a} for roles of the the
  nuclear equation of state in multidimensional hydrodynamic
  modeling. Influences of various neutrino interactions were also
  investigated in \cite{suwa11b, muel12b}.} because it determines the
initial condition and later the accretion rate history. The latter has
a strong leverage on the shock wave evolution, since the force balance
between the ram pressure of preshocked material and the thermal
pressure of postshocked material is the main factor to determine the
shock position and the mass accretion rate, $\dot M=4\pi r^2\rho v$,
is a good measure of the ram pressure, $\rho v^2$. Indeed, it has been
demonstrated that the O-Ne-Mg core of an 8.8 $M_\odot$ star can
produce an explosion even in spherical symmetry thanks to a rapid
decrease in the mass accretion rate onto the shock \citep{kita06}.

Recently, the progenitor dependence of the supernova dynamics is
attracting great attention.
\cite{ugli12} performed a systematic 1D spherically symmetric
simulations for 101 progenitor models from \cite{woos02} with
parametrized neutrino luminosities and demonstrated that the explosion
energy and characteristics of remnant compact objects (neutron stars
and black holes) strongly depend on the initial progenitor structure
\citep[see also][]{ertl15}. Recently, this study was extended by
\cite{naka14} to 2D self-consistent simulations for the same 101
progenitor models.
\cite{ocon13} performed a similar systematic study based on 32
progenitor models from \cite{woos07}, focusing on the compactness
parameter, i.e., the ratio of the mass to radius at a certain mass
coordinate, at the bounce. They found that not the zero age main
sequence (ZAMS) mass but the compactness parameter is a good measure
for the neutrino evolution and hence for shock revival in the
pre-explosion phase.
\cite{couc13c} pointed out that, in addition to the density structure,
velocity fluctuations in progenitors affect the hydrodynamics of shock
revival. They showed in fact that models with velocity fluctuations
imposed before collapse can explode more easily than those without
them. The conclusion was confirmed later by \cite{muel15a}, in which
more systematic parametric studies were performed.  Hence, it is
becoming a consensus in society that the initial condition in
progenitors is an important ingredient in supernova dynamics.

In this paper, we perform a series of neutrino-radiation hydrodynamic
simulations in both spherical symmetry (1D) and axial symmetry (2D)
for progenitors with a mass range from 12 $M_\odot$ to 100 $M_\odot$
in the main sequence phase, and pay attention to the post-bounce
evolutions of mass accretion rates and neutrino luminosities,
particularly how they are affected by the progenitor
structure. Introducing the turning point on the trajectory in the
$\dot M$-$L_\nu$ plane, we argue that its location in the plane will
serve as a sufficient condition for shock revival. We also construct a
phenomenological model to understand how and when the turning point
appears. Applied to a large number of progenitors, the model will be
also useful to judge from the progenitor structure alone which
progenitors are more likely to produce explosions than others.

The paper begins with the descriptions about the numerical simulations
in Section \ref{sec:simulation}. Then, we introduce a new concept,
i.e., the turning point on the trajectory in the $\dot M$-$L_\nu$
plane in Section \ref{sec:critical_turning} and discuss a sufficient
condition for shock revival based on the location of the turning point
relative to the critical curve.  A phenomenological model, which
estimates the mass accretion rate and neutrino luminosity from the
progenitor structure and gives the location of the turning point
qualitatively well, is presented in Section
\ref{sec:phenomenology}. We summarize our results and discuss their
implications in Section \ref{sec:summary}.

\section{Numerical Simulations}
\label{sec:simulation}

\subsection{Methods}
\label{sec:method}

The numerical methods are basically the same we used in our previous
studies \citep{suwa10,suwa11b,suwa13b,suwa14a}. With the ZEUS-2D code
\citep{ston92a} as a base for the hydrodynamics solver, we employ the
equation of state of \cite{latt91} with the incompressibility $K=220$
MeV, for which the maximum mass of a cold NS is 2.04 $M_\odot$, i.e.
more massive than the mass of recently discovered massive NSs
\citep{demo10, anto13}. We solve the neutrino transfer equation for
$\nu_{\e}$ and $\bar\nu_{\e}$ by the isotropic diffusion source
approximation (IDSA) scheme \citep{lieb09} that splits the neutrino
distribution function into two components, which are solved with
different numerical techniques.  The weak interaction rates for
neutrinos are calculated according to \cite{brue85}.
The simulations are performed on a grid of 300 logarithmically spaced
radial zones extending up to 5000 km with the smallest grid width
being 1 km at the center and 128 equidistant angular zones covering
$0\le\theta\le\pi$ for two-dimensional (2D) simulations. For neutrino
transport, we use 20 logarithmically spaced energy bins ranging from 3
to 300 MeV.

We conducted 2D simulations in this paper even though fully 3D
computations with a spectral neutrino transfer are now becoming
possible \citep{taki12,hank13,kuro15,lent15,muel15b}. The main reason
for using 2D simulations is that 3D simulations are computationally
too costly and are not suitable for systematic studies. It is true
that the nature of turbulence is different between 2D and 3D, but its
impact on the critical curve has been disputed by some authors
\citep{murp08,nord10,hank12,couc13b}: 1D certainly gives the highest
critical luminosity but the difference between 2D and 3D is subtle.
It should be noted that the critical curve seems just shifted
vertically with the shape being almost unchanged as the dimension
changes from 1D to 2D to 3D and we hence believe that the
dimensionality will not be a critical factor for the following
arguments in this paper. The treatment of neutrino reactions are not
as sophisticated as in other simulations \citep[e.g.,][]{muel12b} and
heavy neutrinos are neglected in this paper. It is stressed, however,
that our code is calibrated so that some key quantities such as
luminosities of electron-type neutrinos and anti-neutrinos and shock
radii could be reproduced reasonably well in 1D (see next subsection).

We also performed two additional simulations for model s55 employed in
this paper: one with a resolution twice as fine and the other with the
radius of the outer boundary being doubled. In the former computation
we confirmed no remarkable change particularly in neutrino
luminosities whereas in the latter we found that the mass accretion
rate was slightly affected at very late times but the shock evolution
was essentially intact. We hence conclude that the numerical grid
adopted in this paper is adequate at least for the purpose in this
paper.

\subsection{Code check}
\label{sec:check}

In this subsection, we validate our code by comparison with model N13
in \cite{lieb05}, in which the authors performed spherically symmetric
simulations with weak interactions for $\nu_e$ and $\bar\nu_e$ alone
being implemented as described in \cite{brue85} and they demonstrated
that two different codes (AGILE-BOLTZTRAN and VERTEX) produced
consistent numerical solutions. For comparison, we also conduct a
spherically symmetric simulation for the same 13 $M_\odot$ model by
\cite{nomo88}.

\begin{figure}[tbp]
\centering
\includegraphics[width=0.45\textwidth]{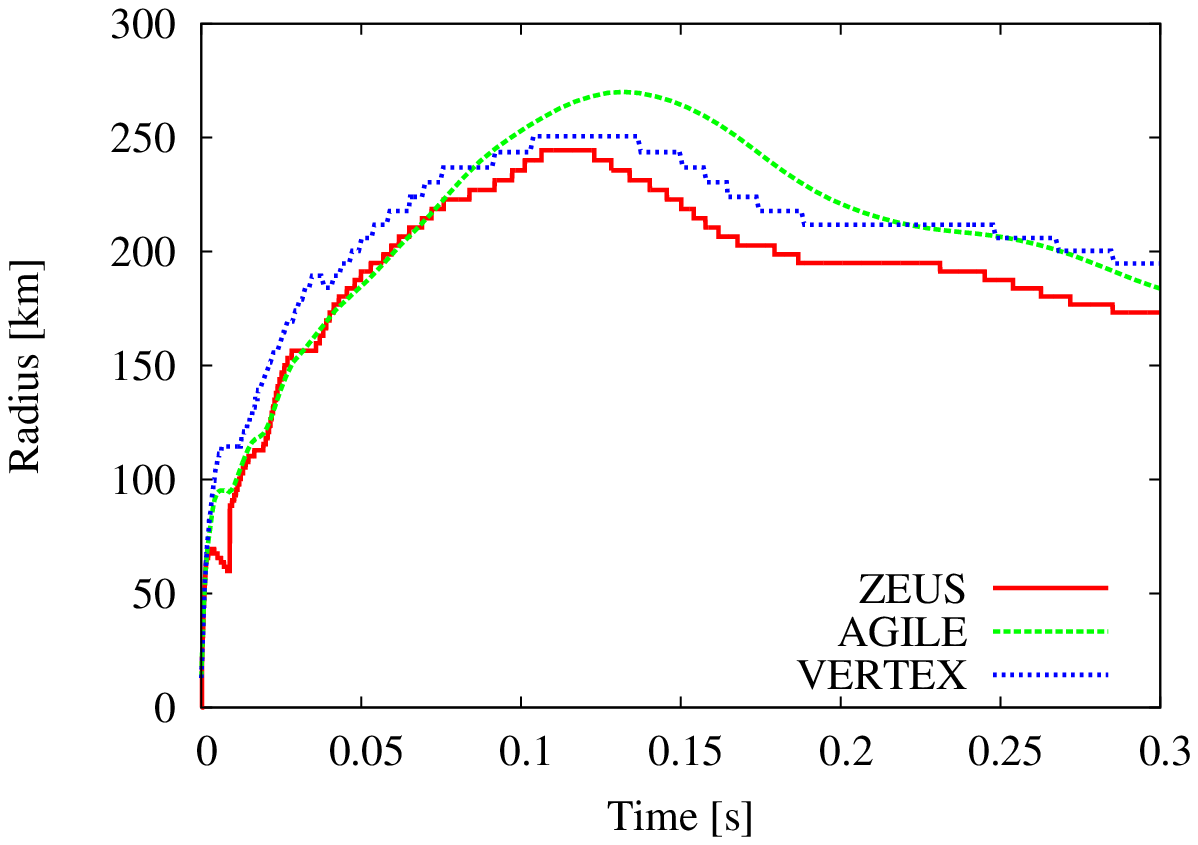}
\includegraphics[width=0.45\textwidth]{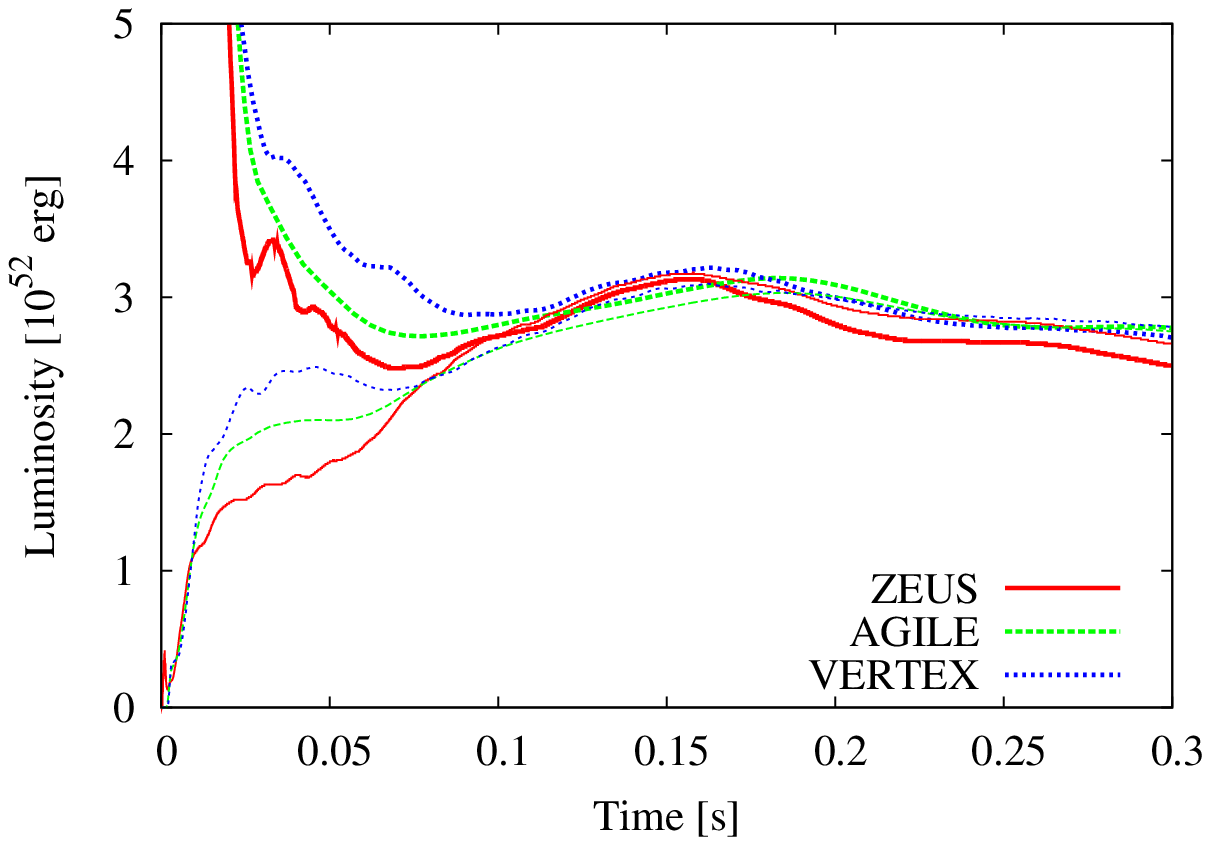}
\caption{Comparison of shock radii (top panel) and neutrino
  luminosities (bottom panel) of ZEUS (code used in this work; red),
  AGILE (\citealt{lieb01}; green), and VERTEX (\citealt{ramp00};
  blue). Luminosities of $\nu_e$ are represented by thick lines and
  those of $\bar \nu_e$ are represented by thin lines.}
\label{fig:comp_lieb05}
\end{figure}

Figure \ref{fig:comp_lieb05} presents the shock radii (top panel) and
luminosities of $\nu_e$ and $\bar\nu_e$ (bottom panel) as a function
of time.  Although small differences can be observed, our results are
consistent with the other two. The slightly smaller shock radii
obtained in our simulation should give more conservative predictions
for shock revival.

\subsection{Progenitor Structures}
\label{sec:progenitor}

We employ progenitors with solar metallicity calculated by
\cite{woos07}, who performed stellar evolutionary calculations. They
have 12, 15, 20, 30, 40, 50, 55, 80 and 100 $M_\odot$ at ZAMS. Some
relevant quantities are presented in Table \ref{tab:progenitor}.

\begin{table}[tbp]
\centering
\caption{Properties of investigated progenitors}
\label{tab:progenitor}
\begin{tabular}{cccccc}
\hline
Model & $M_\mathrm{ZAMS}$ & final mass & final radius & $M_\mathrm{Fe}$ & $R_\mathrm{Fe}$\\
&($M_\odot$) & ($M_\odot$) & ($R_\odot$) &  ($M_\odot$) & (1000 km)\\
\hline
s12  & 12  &    10.91 &   638.41 &    1.285 &    1.061 \\
s15  & 15  &    12.79 &   831.04 &    1.346 &    1.172 \\
s20  & 20  &    15.93 &  1066.68 &    1.540 &    1.591 \\
s30  & 30  &    13.89 &  1552.89 &    1.476 &    1.448 \\
s40  & 40  &    15.34 &    11.80 &    1.804 &    2.123 \\
s50  & 50  &     9.82 &     5.42 &    1.487 &    1.489 \\
s55  & 55  &     9.38 &     0.70 &    1.453 &    1.412 \\
s80  & 80  &     6.37 &     0.60 &    1.479 &    1.501 \\
s100 & 100 &     6.04 &     0.55 &    1.452 &    1.402 \\
\hline
\end{tabular}
\end{table}

\begin{figure}[tbp]
\centering

\subfigure[Density as function of radius]{\includegraphics[width=0.45\textwidth]{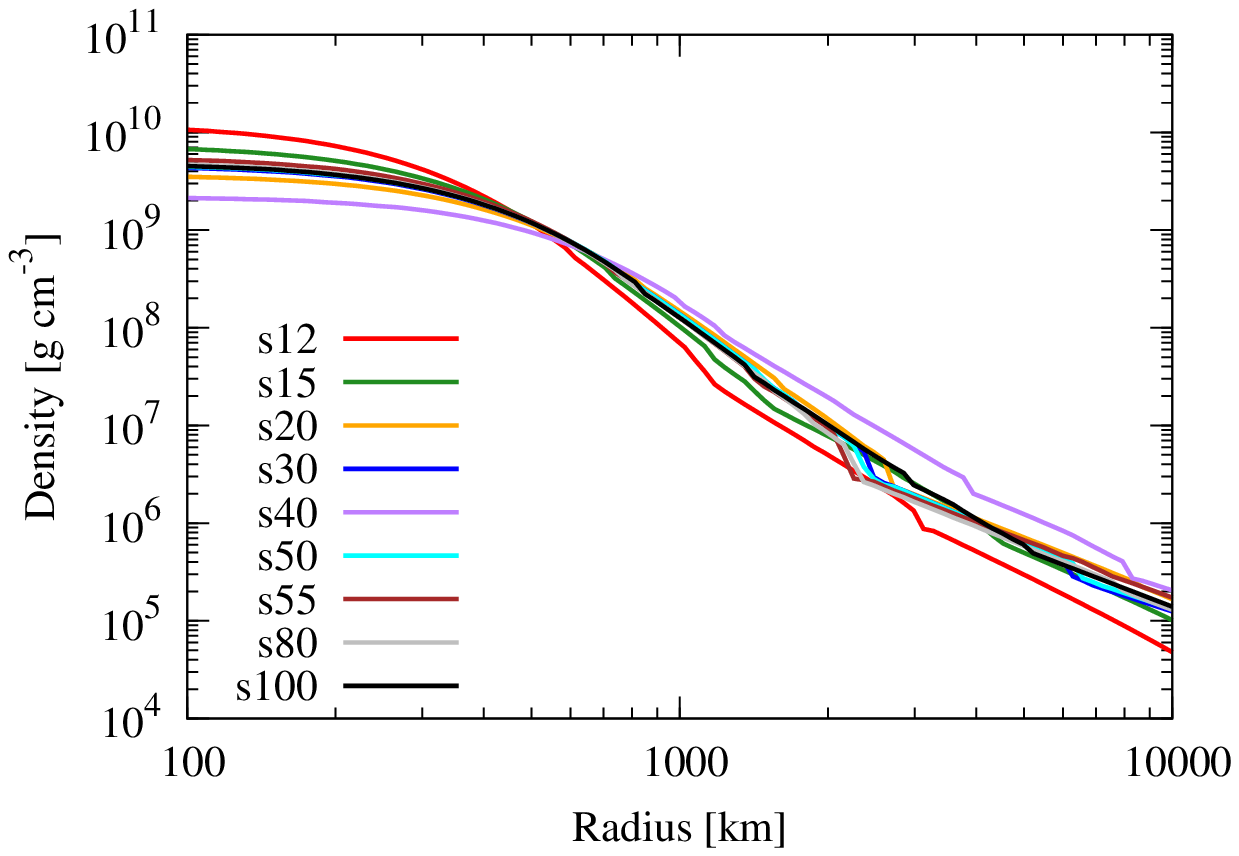}}
\subfigure[Density as function of enclosed mass]{\includegraphics[width=0.45\textwidth]{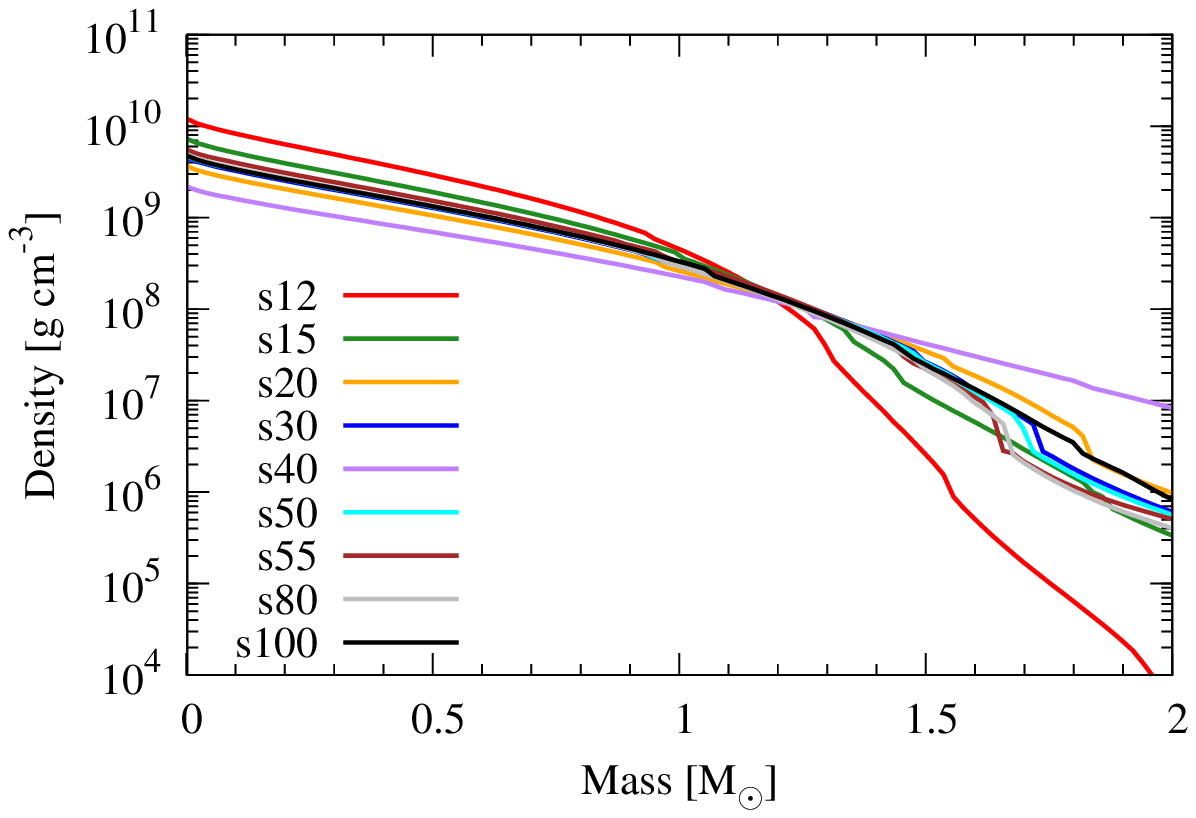}}
\subfigure[Radius as function of enclosed mass]{\includegraphics[width=0.45\textwidth]{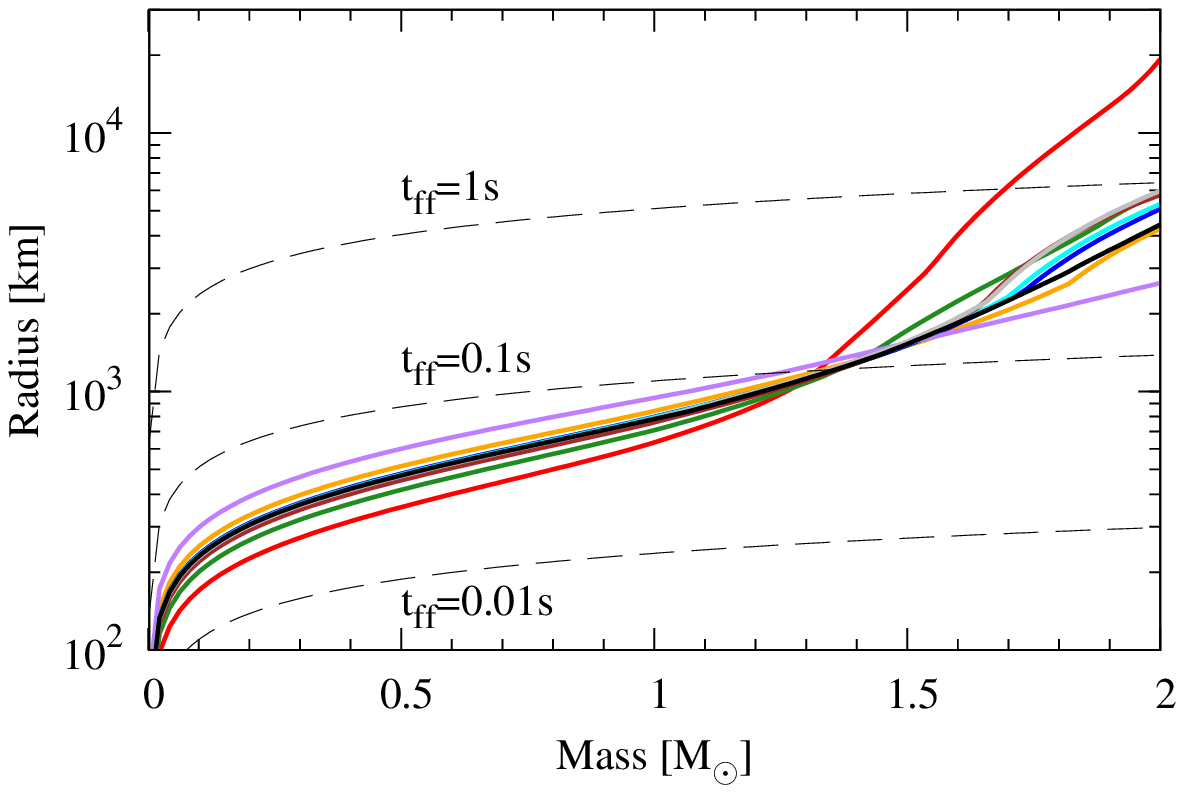}}
\caption{Stellar structures for investigated models. The top two
  panels display the densities as a function of radius (a) and
  enclosed mass (b), respectively. The bottom panel (c) gives the
  radii corresponding to the mass and radius relations. The dashed
  lines show the free-fall times of 0.01, 0.1, and 1 s from bottom to
  top. Refer to text for details.}
\label{fig:density}
\end{figure}

Firstly, we present the structures of these models. Top two panels in
Figure \ref{fig:density} exhibit the density structures as functions
of the radius (panel (a)) and enclosed mass (panel (b)). Panel (c)
shows the mass-radius relation, in which the free-fall timescales
($t_\mathrm{ff}=\sqrt{r^3/GM}$, where $r$ is the radius, $G$ is the
gravitational constant, and $M$ is the enclosed mass) are plotted as
dashed lines.  One can find that the density structure and ZAMS mass
do not correlated with each other in a simple way: the density at the
enclosed mass of 2 $M_{\odot}$ (see panel (b)) becomes the smallest
for the model with a ZAMS mass of 12 $M_{\odot}$ and attains the
maximum at 40 $M_{\odot}$. Models with the ZAMS masses larger than 40
$M_{\odot}$ have densities in between. This is because strong mass
loss during the main sequence and giant phases yields smaller cores
(see also Table \ref{tab:progenitor}). From dashed lines in panel (c),
one can easily see that the difference of structure leads to different
free-fall time of mass elements, which will then result in different
mass accretion histories.

\begin{figure}[tbp]
\centering
\includegraphics[width=0.45\textwidth]{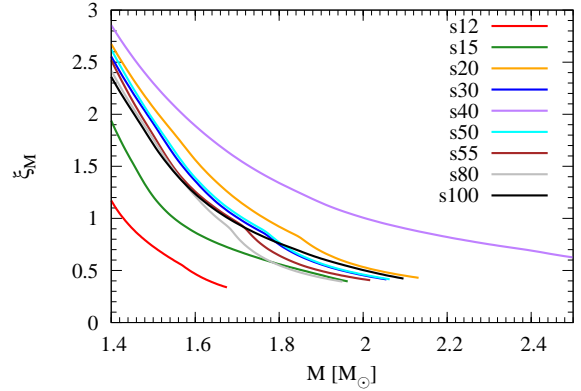}
\caption{The {\it compactness} parameters $\xi_{M}$ defined in
  Eq. (\ref{eq:xi_M}) as a function of mass coordinate $M$. A lager
  $\xi_{M}$ means a more compact structure: s12 is the least compact
  progenitor, while s40 is the most compact.}
\label{fig:xi}
\end{figure}

In Figure \ref{fig:xi}, we show the compactness parameter defined by
\cite{ocon11} as
\begin{equation}
\xi_{M}=\left.\frac{M/M_{\odot}}{R(M)/1000~\mathrm{km}}\right|_{t=t_\mathrm{bounce}},
\label{eq:xi_M}
\end{equation}
where $R(M)$ denotes the radius for the enclosed mass $M$. It is
evident that the 12 $M_{\odot}$ model has the smallest $\xi_{M}$ at
all enclosed masses. $\xi_{M}$ increases with the progenitor mass up
to 40 $M_{\odot}$. Interestingly, the models with 50, 55, 80 and 100
$M_{\odot}$ have smaller $\xi_{M}$ than the model with 40 $M_{\odot}$,
which is consistent with the results of \cite{ocon13}, in which they
showed that the model with 40 $M_\odot$ gives the maximum values both
for $\xi_{1.75}$ and $\xi_{2.5}$. In this sense the model s40 is the
most compact progenitor, whereas the model s12 is the least compact
one. Recently, the temporal evolutions of the $\xi$ parameter during
the stellar evolution were studied in detail by \cite{sukh14} and it
was shown that $\xi$ is indeed an appropriate quantity to characterize
the progenitor structure.

The subsequent subsections present our numerical simulations in 1D
and 2D consecutively.

\subsection{Spherically symmetric simulations}
\label{sec:1D}

\begin{figure}[tbp]
\centering
\subfigure[Shock radius evolution]{\includegraphics[width=0.45\textwidth]{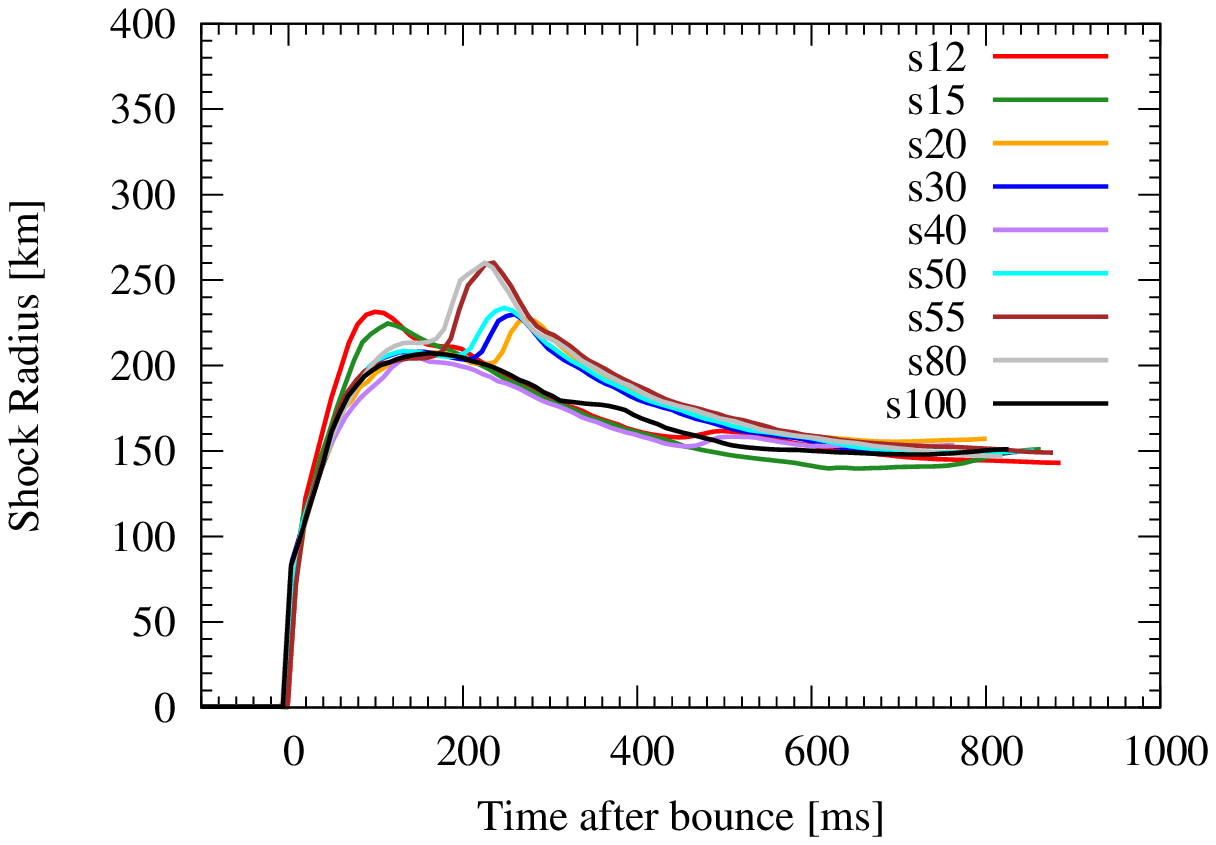}\label{fig:1d_a}}
\subfigure[Mass accretion rate evolution]{\includegraphics[width=0.45\textwidth]{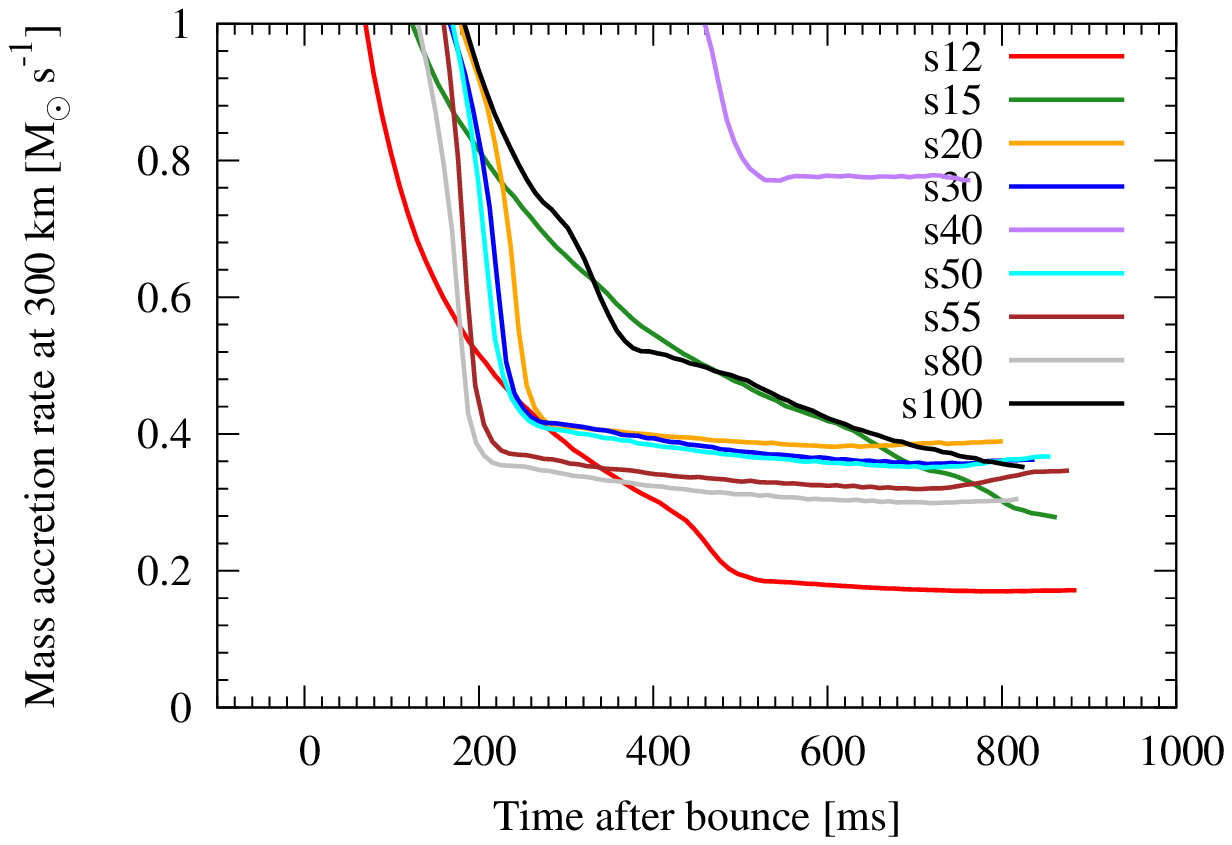}\label{fig:1d_b}}
\caption{The time evolutions of shock radius (a) and mass accretion
  rate (b). There are bumps in panel (a), which correspond to the
  rapid decreases of mass accretion rate (see panel (b)).}
\label{fig:1d}
\end{figure}

In Figure \ref{fig:1d}, we show the time evolutions of shock radius
(panel (a)) and mass accretion rate at 300 km (panel (b)). It can be
seen that all simulations failed to explode due to insufficient
neutrino heating just as expected. We especially pay attention to the
evolutions of shock radius and the mass accretion rate, which are
intimately related with each other.

The lower panel in the Figure \ref{fig:1d} shows that the mass
accretion rate decreases rapidly until $\sim$200 ms and then becomes
almost constant thereafter for a majority of models including s20,
s30, s50, s55, and s80. This transition in the mass accretion rate
originates from the change of accreting layers, i.e., from the silicon
layer to the oxygen layer: normally, a large density jump at the
boundary of these layers in progenitors can be observed
\cite[see][]{woos07}. As a consequence of this transition, the stalled
shock wave expands for a while around the sudden change in the mass
accretion rate.
We can also recognize some exceptions: model s12 shows a similar
change in the mass accretion rate but at a much later time $\sim$500
ms with a smaller mass accretion rate; in model s100 the mass
accretion rate is not settled to a constant value after the change at
$\sim$350 ms but continues to decrease thereafter; in model s15 there
is no clear change discernible at all. In the former two cases, the
shock expands slightly or stops receding for a while around the times
of the changes in the mass accretion rate like the first group. Model
s40 is another outlier; although it has a similar pattern in the
temporal evolution of the mass accretion rate to those found for the
majority of progenitors, the transition occurs at a much later time
$\sim$500 ms with a much high accretion rate. As a result of this high
mass accretion and because of a larger ram pressure, the temporary
expansion of the stalled shock wave associated with the transition is
much less remarkable.

\begin{figure*}[tbp]
\centering
\includegraphics[width=0.7\textwidth]{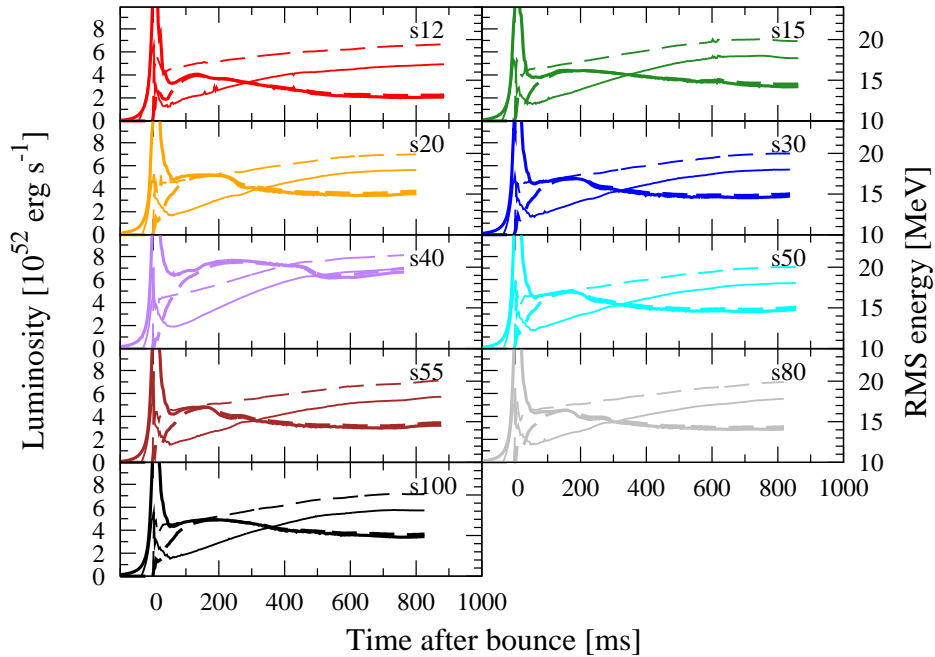}
\caption{Neutrino luminosities (thick lines) and RMS energy (thin
  lines). Solid lines represent $\nu_e$ and dashed lines give
  $\bar\nu_e$.}
\label{fig:neutrino}
\end{figure*}

In Figure \ref{fig:neutrino}, we show time evolution of the neutrino
luminosity $L_\nu$ (thick lines) and RMS energy (thin lines), which is
determined by $\sqrt{\bracket{\epsilon_\nu^2}}=\sqrt{(\int
  d\epsilon_\nu d\mu f_\nu \epsilon_\nu^5)/(\int d\epsilon_\nu d\mu
  f_\nu \epsilon_\nu^3)}$, where $\epsilon_\nu$ is the neutrino
energy, $\mu$ is the cosine of angle between radial and neutrino
propagation directions, and $f_\nu$ is the distribution function of
neutrinos, for $\nu_e$ (solid lines) and $\bar\nu_e$ (dashed
lines). Note that the neutrino heating rate can be written as
$Q_\nu^+\propto L_\nu\bracket{\epsilon_\nu^2}/r^2$. In spite of the
difference of mass accretion history (see Figure \ref{fig:1d_b}), the
neutrino luminosities and RMS energies are not very different. Among
them, model s12 and model s40 exhibit the smallest and the highest
values, respectively.

\begin{figure}[tbp]
\centering
\includegraphics[width=0.45\textwidth]{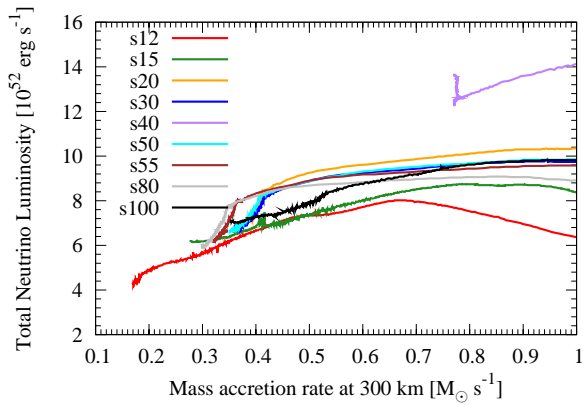}
\caption{Model trajectories in the $\dot M$-$L_{\nu}$ plane. The mass
  accretion rate is evaluated at 300 km from the center.}
\label{fig:mdot-lnu}
\end{figure}

In Figure \ref{fig:mdot-lnu}, we show the same evolutions in the $\dot
M$-$L_\nu$ plane, where $\dot M$ is the mass-accretion rate at 300 km
and $L_\nu$ is the total neutrino luminosity (i.e. the sum of the
contributions from $\nu_e$ and $\bar{\nu}_e$).  Each model moves from
right (high-accretion rates) to left (low-accretion rates) on these
curves. As the mass accretion rate decreases, the neutrino luminosity
also diminishes in general except possibly for the very early phase.
One can recognize the steepening of these curves near their left ends
for the same majority group of progenitors. This is yet another
manifestation of the transition in the mass accretion rate in this
plane: the mass accretion rate becomes almost constant after the
transition while the neutrino luminosity continues to decrease.  The
position of this transition point in the $\dot M$-$L_\nu$ plane is
important for shock revival, which will be discussed in detail in the
next section. If the point is located more to the top left corner in
this plane (i.e., having lower mass accretion rates and higher
neutrino luminosities), such a model will be more likely to produce an
explosion, particularly in multi-dimensional simulations, in which the
critical curve is supposed to run lower than in 1D. Models s55 and s80
are hence good candidates for exploding models in 2D in this sense.

\begin{figure}[tbp]
\centering
\includegraphics[width=0.45\textwidth]{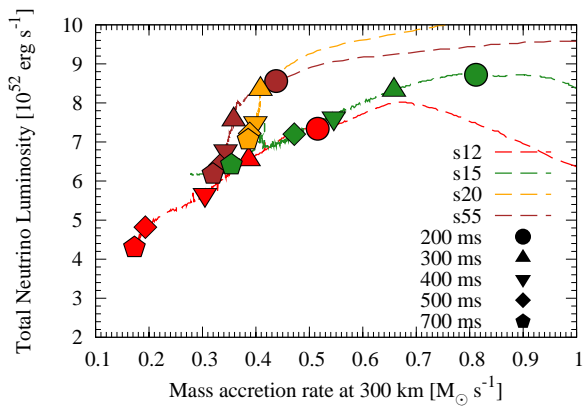}
\caption{Model trajectories in the $\dot M$-$L_{\nu}$ plane for
  selected models with various points representing the times after
  bounce. The colors of lines are the same as those in Figure
  \ref{fig:mdot-lnu}. Each model evolves from right (high accretion
  rate) to left (low accretion rate).}
\label{fig:mdot-lnu_limited}
\end{figure}

Figure \ref{fig:mdot-lnu_limited} is the same as Figure
\ref{fig:mdot-lnu} but for selected models, i.e., s12, s15, s20, and
s55, with some time stamps. The other models not shown in Figure
\ref{fig:mdot-lnu_limited} have similar trajectories. Models s20 and
s55, members of the majority group, have a clear transition point ---
will be referred to as the {\it turning point} in the next section ---
at $(\dot M, L_\nu)\approx(0.4M_\odot~\mathrm{s^{-1}}, 9\times
10^{52}~\mathrm{erg~s}^{-1})$ and
$\approx(0.35M_\odot~\mathrm{s^{-1}}, 8\times
10^{52}~\mathrm{erg~s}^{-1})$, respectively. This is due to rather
large jumps in density at the boundary between the silicon and oxygen
layers in these models, which are apparent in Figure
\ref{fig:density}.  As it is also evident from the time stamps in
Figure \ref{fig:mdot-lnu_limited}, these models move rapidly from
right (high accretion rate) to left (low accretion rate) up to the
turning point and then shift downwards slowly later. They hence stay
near the turning point for a long time and, as argued later, that is
the point, where shock revival is most likely to occur. This is why we
propose to employ the position of the tuning point as a diagnostic of
explosion.  It is also noted that the turning point is not always
clearly visible and may not exist at all for some models. As mentioned
earlier, model s12 does show a transition in the mass accretion rate
in Figure \ref{fig:1d_b} but the turning point is barely visible near
the left end of the trajectory. Model s15 does not show any
discernible transition already in Figure \ref{fig:1d_b}, which is also
reflected in Figure \ref{fig:mdot-lnu_limited}. It is hence clear that
the criterion for shock revival later proposed is a sufficient
condition applicable only to those cases with the turning point.

\subsection{Axially symmetric simulations}
\label{sec:2D}

In this subsection, we show the results of 2D simulations for the
progenitors explored in the previous subsection in 1D. Since all 1D
models fail to explode, these 2D simulations will serve as a guide to
consider which progenitors are likely to explode.

\begin{figure*}[tbp]
\centering
\includegraphics[width=0.49\textwidth]{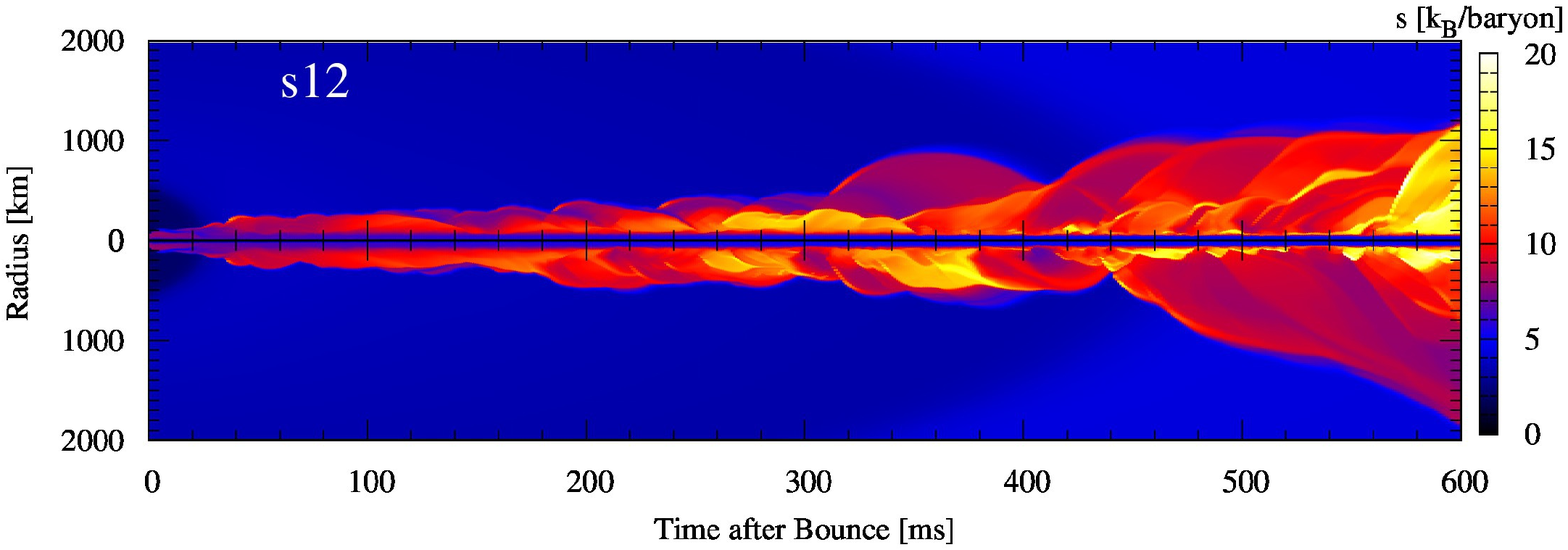}
\includegraphics[width=0.49\textwidth]{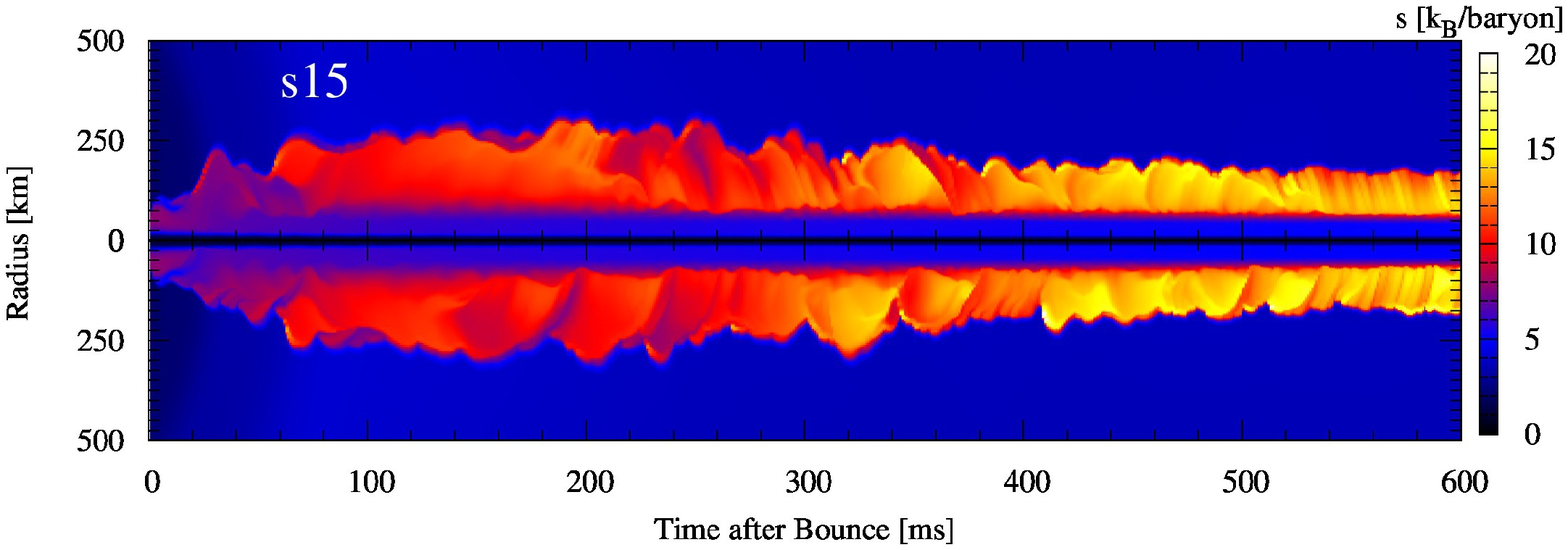}
\includegraphics[width=0.49\textwidth]{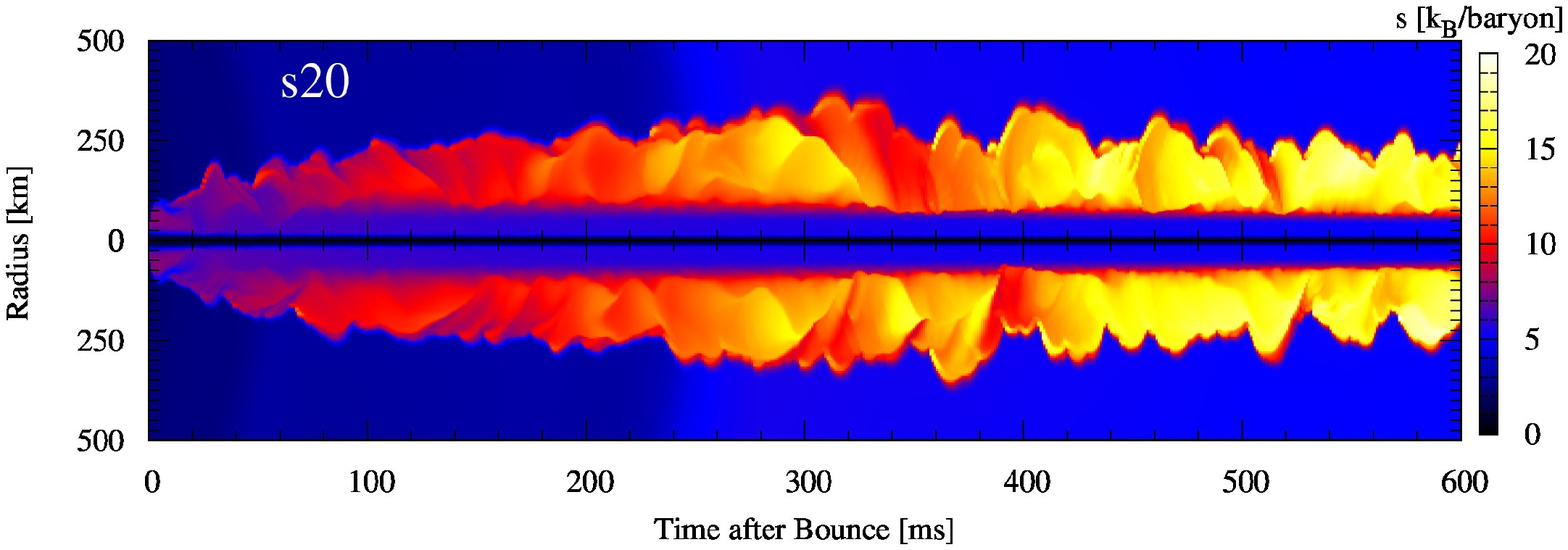}
\includegraphics[width=0.49\textwidth]{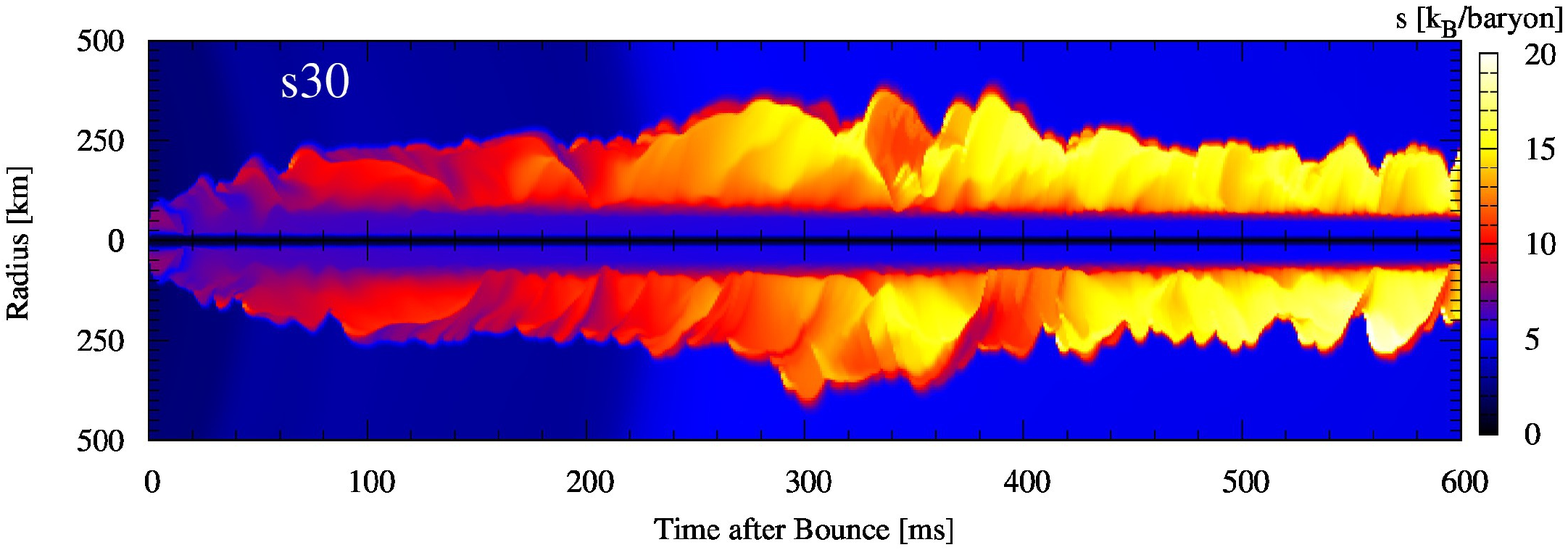}
\includegraphics[width=0.49\textwidth]{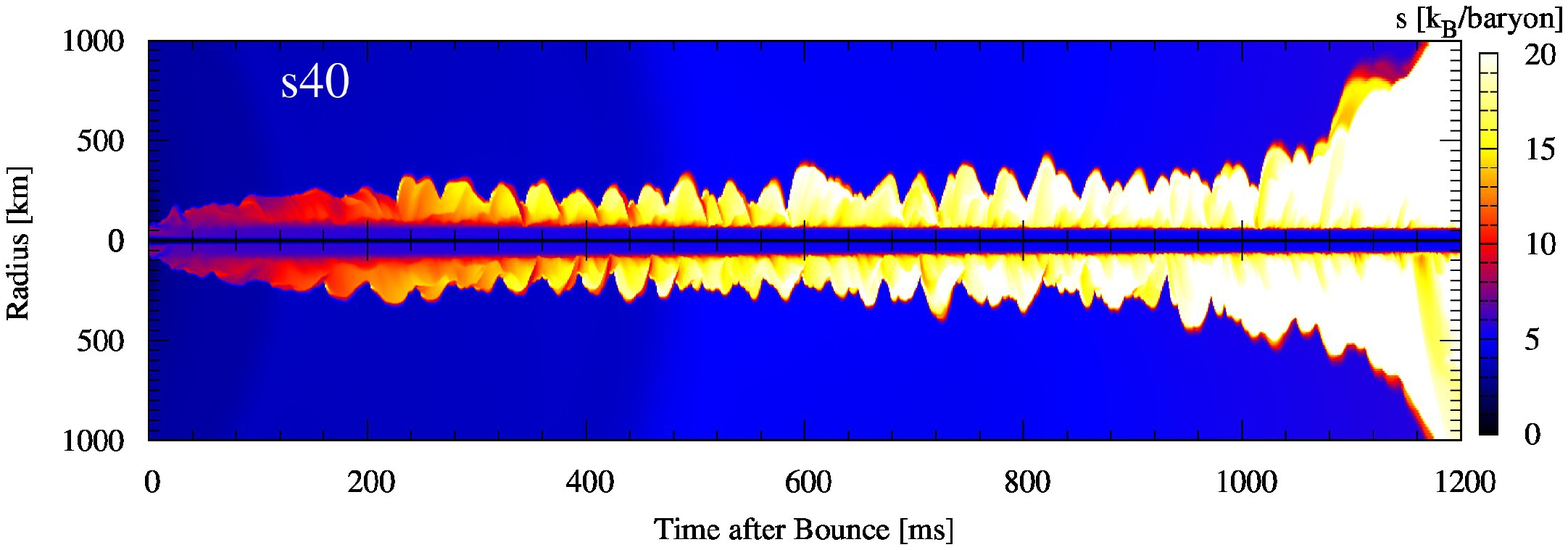}
\includegraphics[width=0.49\textwidth]{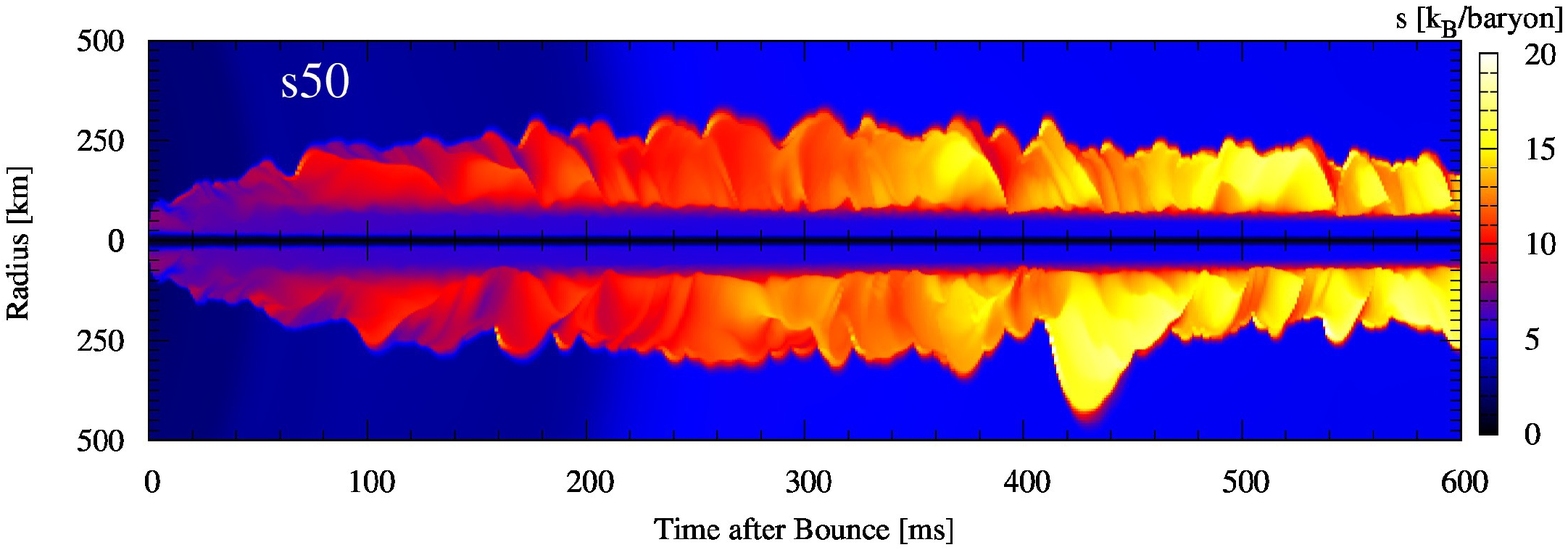}
\includegraphics[width=0.49\textwidth]{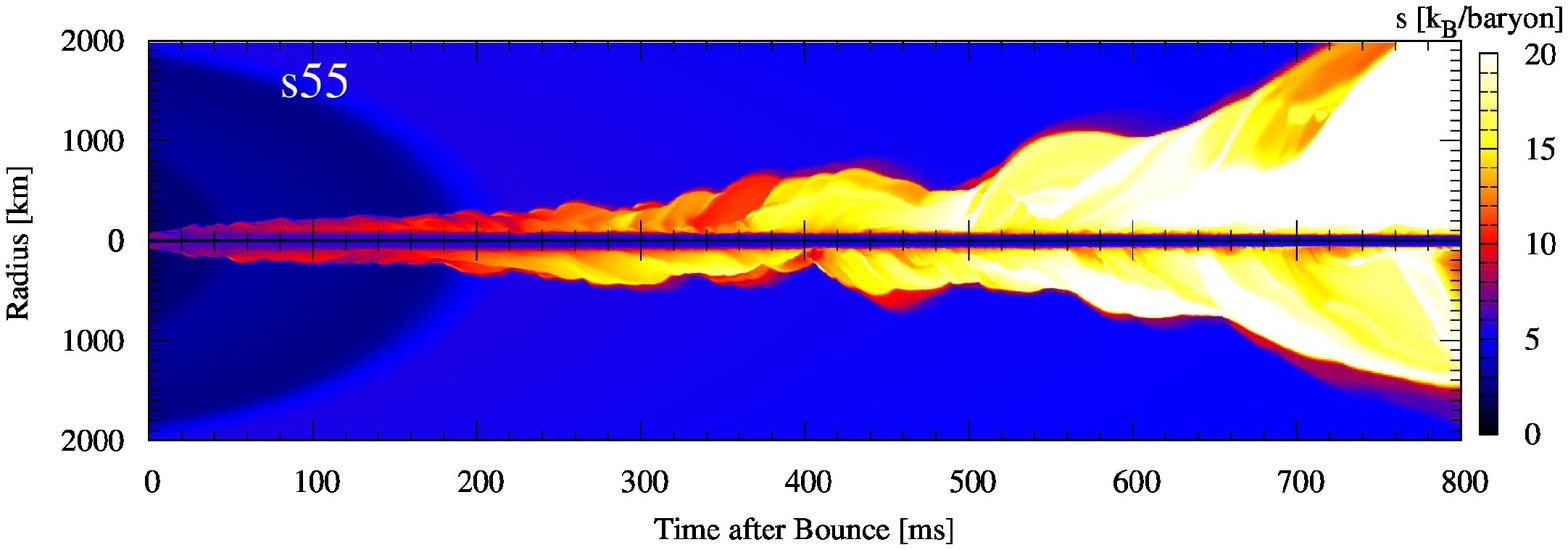}
\includegraphics[width=0.49\textwidth]{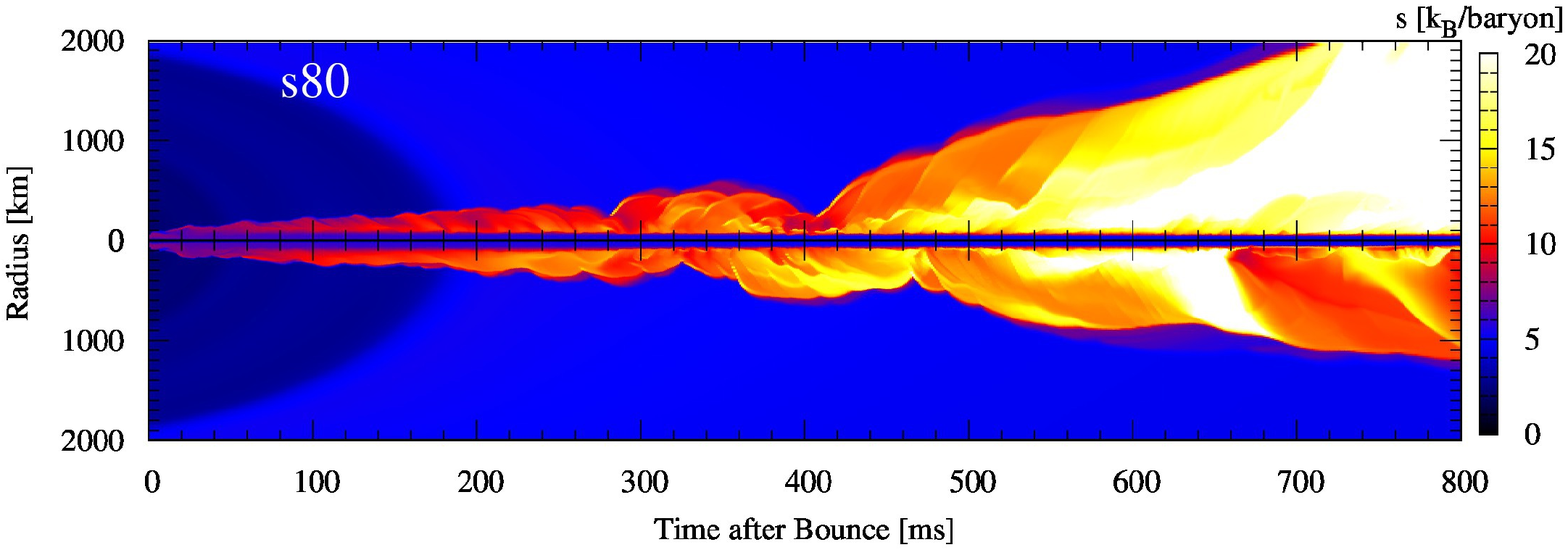}
\includegraphics[width=0.49\textwidth]{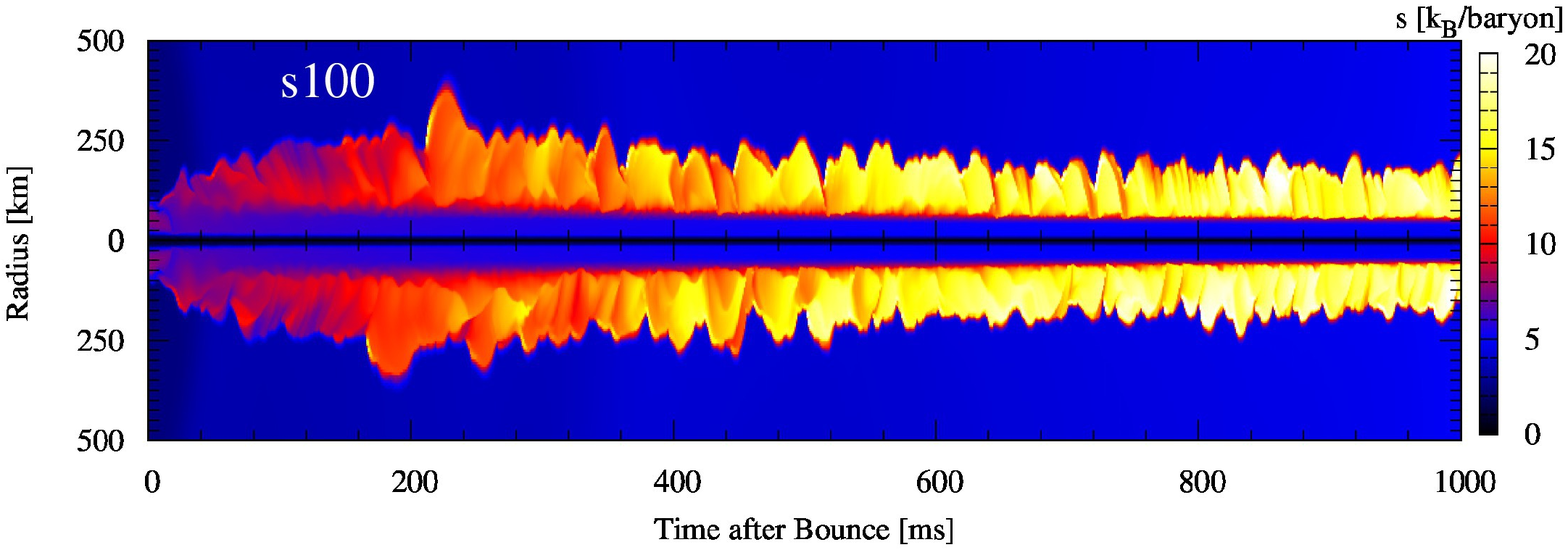}
\caption{Time-space diagrams of specific entropy at poles for
  two-dimensional simulations. Upper (lower) panels represent the
  values at the north (south) pole. Models s12, s40, s55, and s80
  eventually produce explosions at different times, depending on the
  initial density structures. The other progenitors, i.e., s15, s20,
  s30, s50, and s100, failed to produce an explosion at least by the
  end of simulations.}
\label{fig:entropy_pole}
\end{figure*}

Figure \ref{fig:entropy_pole} gives the temporal evolutions of entropy
at the north (top panels) and south poles (bottom panels),
respectively. There are several oscillations in the shock radius for
these models, which are the consequences of the standing accretion
shock instability (SASI)
\citep{blon03,ohni06,blon07,fogl07,iwak08,fern09}. It is clear that
the material in the postshock region is heated up by neutrino
irradiation from PNS (the yellow color represents high
entropies). Thanks to this heating, some models (s12, s40, s55, and
s80) eventually produce shock re-expansion along the symmetry
axis. The other models (s15, s20, s30, s50, and s100) yield no such
expansion at least by the end of simulations even though there is
certainly neutrino heating in operation.

\begin{figure}[tbp]
\centering
\includegraphics[width=0.45\textwidth]{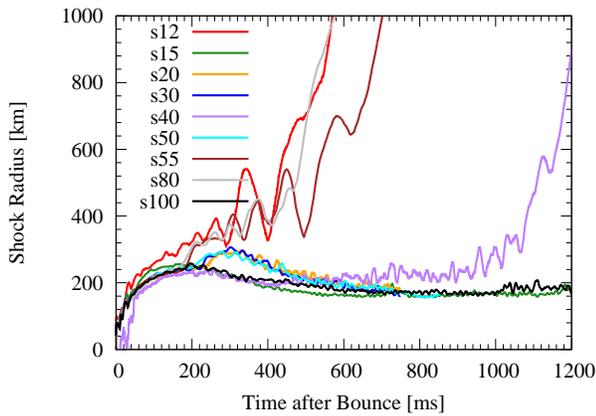}
\caption{Time evolutions of the angle averaged shock wave radius. Four
  of the investigated models, i.e. s12, s40, s55 and s80, clearly show
  shock expansions.}
\label{fig:shock_2D}
\end{figure}

Figure \ref{fig:shock_2D} presents time evolutions of the shock radius
averaged over a solid angle. It can also be seen in this figure that
several progenitors produce a shock revival, which is a necessary
condition for a supernova explosion.\footnote{Note that a shock
  revival, or a re-expansion of the stalled shock wave, is not a
  sufficient condition for a supernova explosion. In fact, shock
  revival is just a consequence of the dominance of the post-shock
  thermal pressure over the ram pressure in the pre-shocked region and
  the mass accretion to PNS may continue thereafter, increasing the
  mass of PNS. In order to produce a {\it successful} explosion, the
  expanding shock should be strong enough to turn the accretion to an
  expansion of the envelope. See \cite{suwa13b} for more details.}  It
is important to note that these successful models have the turning
points located either more to the left (s12; low accretion rate) or
more to the top (s40; high neutrino luminosity) or both (s55, s80; low
accretion rate and high luminosity) than unsuccessful models. We use
this observation in the next section.

It seems that the onset of shock revival is delayed from the time of
the turning point (see Figure \ref{fig:1d_b}). This is because the
mass accretion rates in Figure \ref{fig:1d_b} are evaluated at 300 km
from the center and it takes some time until it influences the post
shock dynamics. Note also that the development of shock oscillations
needs some time.
It should be mentioned, however, that s40 will probably fail to
explode when we take into account general relativity, which is
neglected in this paper. \cite{ocon11} observed in their 1.5D general
relativistic simulation that the same progenitor formed a black hole
at around 550 ms after bounce (similar results were obtained by
\citealt{sumi06} and \citealt{fisc09} but with different progenitor).
Since this time of black hole formation is much earlier than the shock
revival time we found in s40, the progenitor leads most likely to a BH
formation instead of the very late explosion observed here. Note,
however, that s40 is an outlier anyway, having a very large
compactness and a very late occurrence of the turning point.

\begin{figure}[tbp]
\centering
\subfigure[Abundance distribution and density structure]{\includegraphics[width=0.45\textwidth]{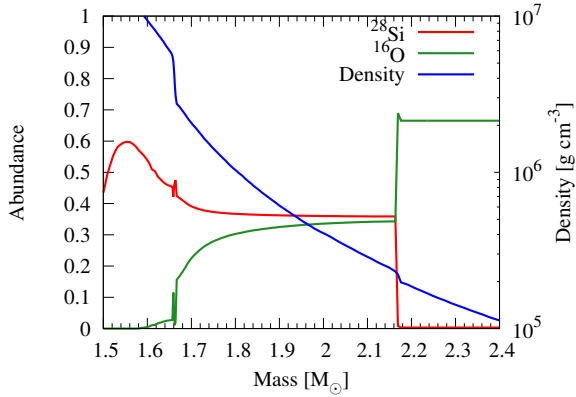}}
\subfigure[Time evolution of mass coordinate and shock]{\includegraphics[width=0.45\textwidth]{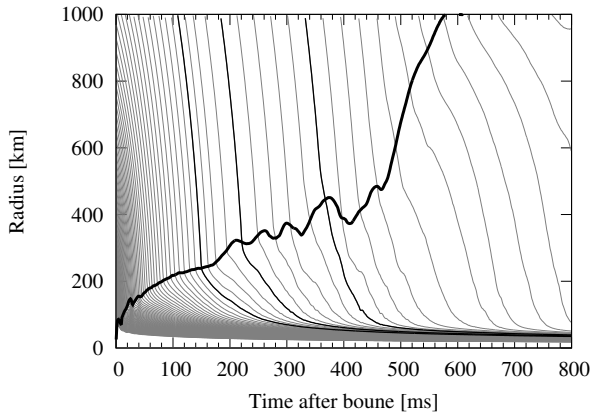}}
\caption{(Top) The initial profiles of density and composition for
  model s80. The abundance of $^{28}$Si (red line) and $^{16}$O (green
  line), and the density (blue line) are given as a function of the
  mass coordinate. There are two jumps in density, representing the
  transition of layers.  (Bottom) Trajectories of the mass shells with
  the mass coordinates of 1 $M_{\odot}$ to 1.85 $M_{\odot}$ with an
  interval of 0.01 $M_{\odot}$ are plotted as grey curves for the same
  model. Thin black lines represent 1.66, 1.7, and 1.75 $M_\odot$ from
  left to right, respectively. A thick black curve indicates the
  average shock position. When the mass shell of 1.66 $M_{\odot}$ runs
  across the shock several oscillations ensue in the shock radius. The
  shock is eventually expanded at $\sim 400$ ms after the bounce.}
\label{fig:ab_den}
\end{figure}

The top panel of Figure \ref{fig:ab_den} exhibits the abundance of
$^{28}$Si (red line) and $^{16}$O (green line) as well as the density
(blue line) of model s80. One can find that there are two density
jumps at 1.66 $M_{\odot}$ and 2.17 $M_{\odot}$ in mass coordinate. The
bottom panel of this figure displays as gray lines the trajectories of
mass shells at the mass coordinates of 1 $M_{\odot}$ to 1.85
$M_{\odot}$ with an interval of 0.01 $M_{\odot}$. Three thin black
lines represent the mass coordinates of 1.66, 1.7, and 1.75
$M_{\odot}$. Note that 1.66 $M_{\odot}$ corresponds to the interface
between Si and oxygen burning shells (see also panel (a)). It is
interesting to see what happens when this mass shell accretes onto the
shock (thick black line). It is evident that several oscillations
ensue and the standing shock is finally converted to the expanding
shock at $\sim 400$ ms after the bounce. This is a clear demonstration
that the transition in the mass accretion rate triggers shock revival.

\begin{figure}[tbp]
\centering
\includegraphics[width=0.45\textwidth]{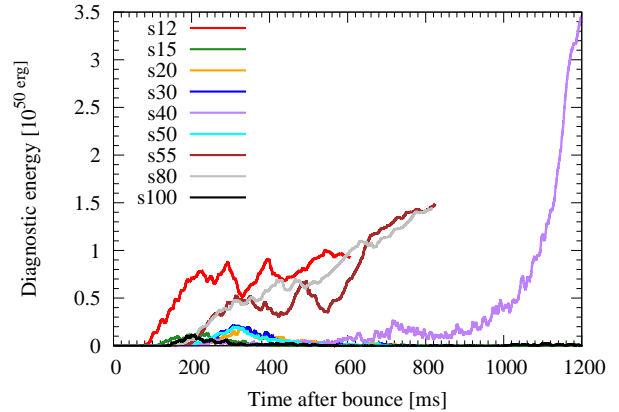}
\caption{Time evolutions of the diagnostic energy for 2D models. It is
  defined by the integral of the sum of the specific internal,
  kinetic, and gravitational energies, over the zones, in which it is
  positive. The horizontal axis is the postbounce time.}
\label{fig:diagnostic_energy}
\end{figure}

Although this is not relevant for the main focus of this paper, we
show in Figure \ref{fig:diagnostic_energy} for reference the so-called
diagnostic energy, which is defined as the integral of the sum of
specific internal, kinetic and gravitational energies over all zones,
in which it is positive.  Four exploding models (s12, s40, s55 and
s80) have indeed non-vanishing diagnostic energies. Some oscillations
originate from the shock oscillations. Though the diagnostic energy is
gradually increasing, the final value is still much smaller than the
typical value of the observed explosion energy, $\sim 10^{51}$
erg. Although even non-exploding models have positive diagnostic
energies due to neutrino heating, it is insufficient to revive the
stalled shock wave.

\subsection{Different $15 M_\odot$ models}

\begin{figure}[tbp]
\centering
\subfigure[Density as a function of radius]{\includegraphics[width=0.45\textwidth]{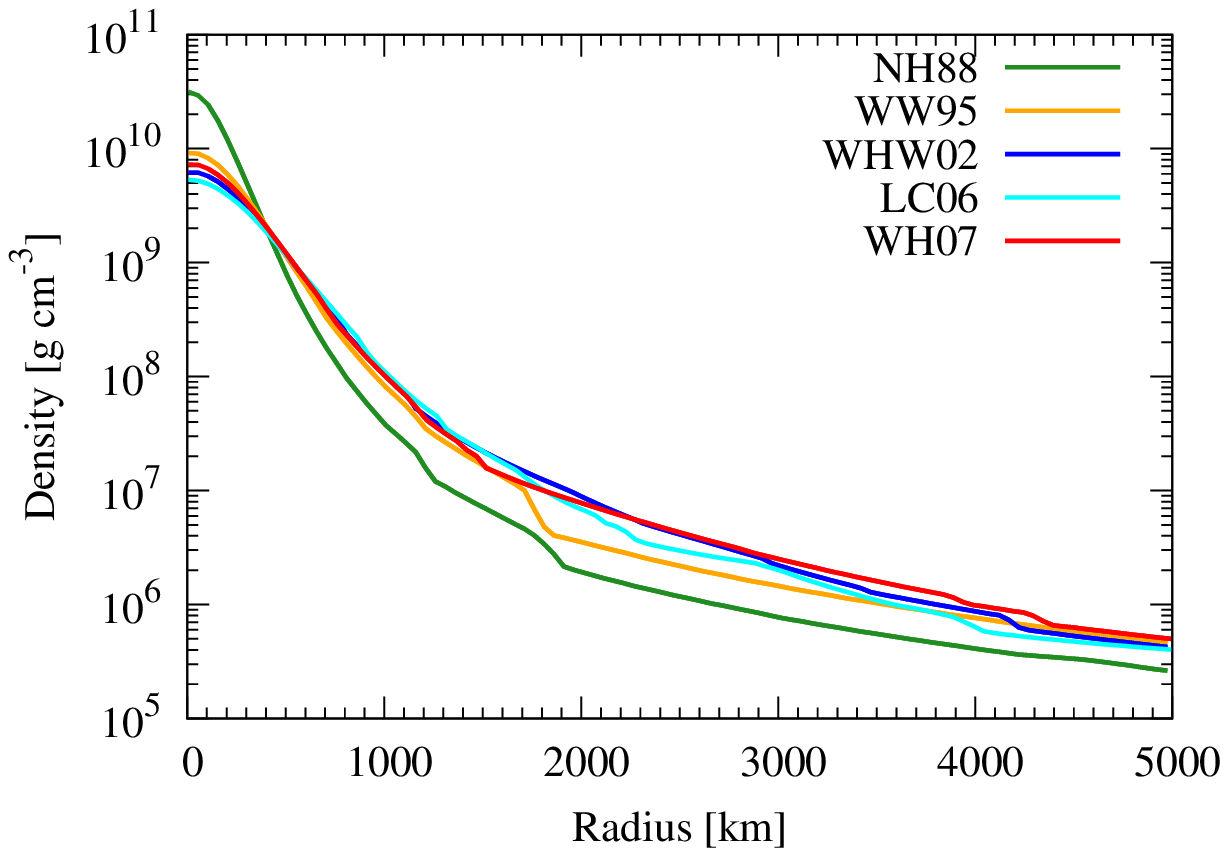}}
\subfigure[Radius as a function of enclosed mass]{\includegraphics[width=0.45\textwidth]{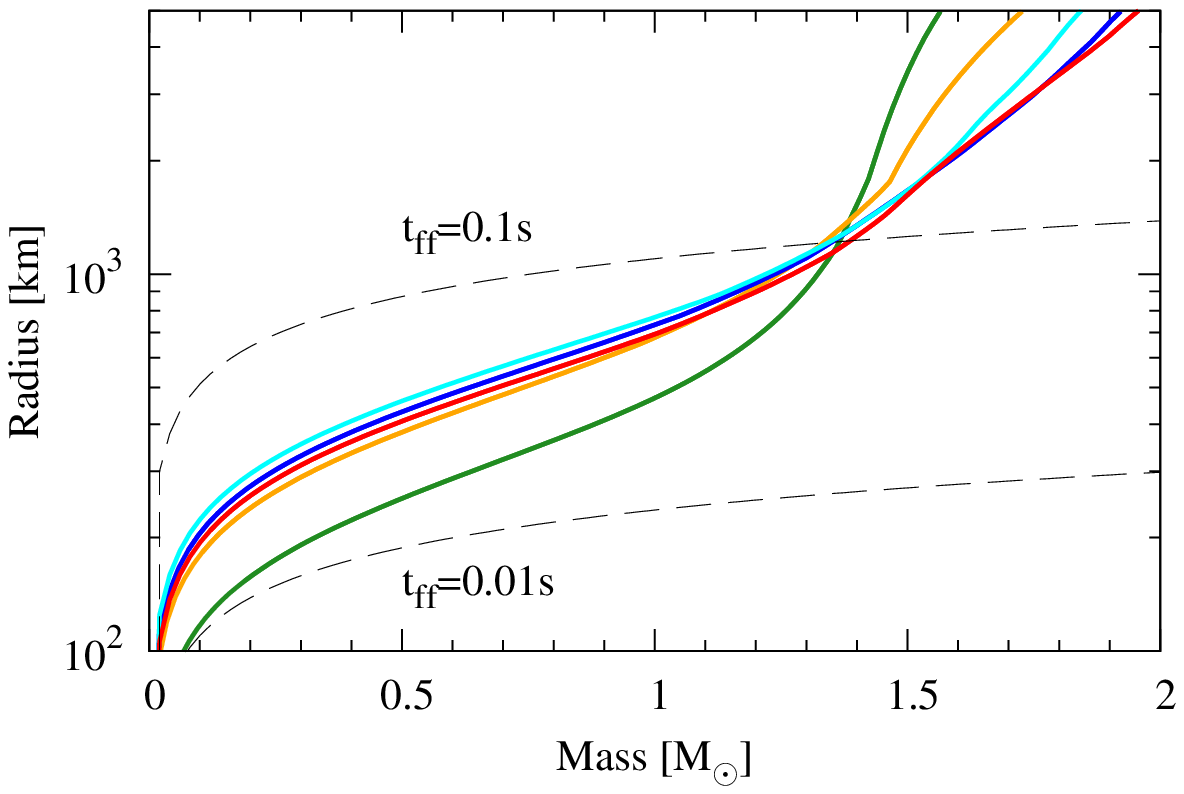}}
\caption{The same as Figure \ref{fig:density} but for progenitors with
  the ZAMS mass of 15 $M_\odot$. Here we use five models from
  \cite{nomo88} (NH88), \cite{woos95} (WW95), \cite{woos02} (WHW02),
  \cite{limo06} (LC06), and \cite{woos07} (WH07). Due to the different
  treatments of physics and numerics for stellar evolutionary
  calculations, the structures prior to collapse show diversity even
  if they have the same ZAMS mass. In the bottom panel, free-fall
  times are given by dashed lines.}
\label{fig:density_15}
\end{figure}

It is a well known unfortunate fact that stellar evolution
calculations by different groups do not agree with one another.
In this subsection, we investigate this issue, employing different
progenitors with the same typical mass of $15 M_\odot$ at ZAMS. In
addition to model s15 just studied, we use four more models from
\cite{nomo88} (NH88), \cite{woos95} (WW95), \cite{woos02} (WHW02), and
\cite{limo06} (LC06).  The first three of them were also employed in
\cite{suwa11b}, in which neutrino oscillation effects on a supernova
explosion were investigated. The pre-collapse density structures are
given in Figure \ref{fig:density_15} (see also Figure 8 of
\citealt{suwa11b} for comparison of the density structures at 100 ms
after the bounce. In this paper it was argued that the structures are
similar among the different models for $M<0.8 M_\odot$ whereas they
are different for $M>0.8M_\odot$). It can be observed that even though
the initial mass at ZAMS is the same, the density structures prior to
collapse become different, depending on both the physics and the
numerics implemented in stellar evolutionary calculations.  It should
be noted in particular that the difference between WW95 and WH07 is
substantial for $M\gtrsim 1.1M_\odot$ before collapse (compare red and
orange lines in Figure \ref{fig:density_15}).

\begin{figure}[tbp]
\centering
\includegraphics[width=0.45\textwidth]{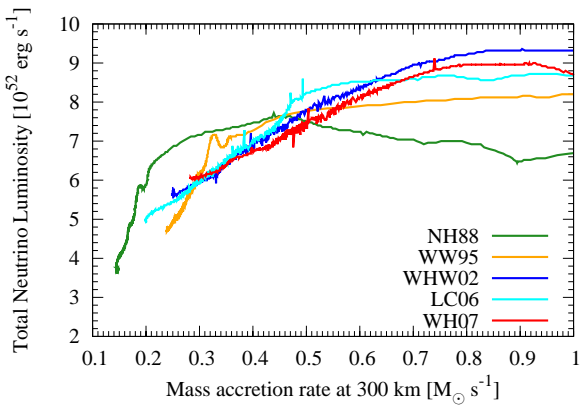}
\caption{Model trajectories in the $\dot M-L_\nu$ plane for the 1D
  simulations of 15 $M_\odot$ progenitors. This is the same as Figure
  \ref{fig:mdot-lnu} but for different progenitor models. The mass
  accretion rate is evaluated at 300 km from the center. }
\label{fig:mdot-lnu_15}
\end{figure}

Figure \ref{fig:mdot-lnu_15} presents these models in the $\dot
M-L_\nu$ plane evaluated for 1D simulations (cf. Figure
\ref{fig:mdot-lnu}). NH88, WW95 and LC06 have clear turning points and
that the former two are located more to the left than the last and are
more likely to achieve shock revival.  This is a consequence of the
density jumps more remarkable for these models as observed in Figure
\ref{fig:density_15}. It is noted that all 1D simulations failed to
produce an explosion.

\begin{figure}[tbp]
\centering
\includegraphics[width=0.45\textwidth]{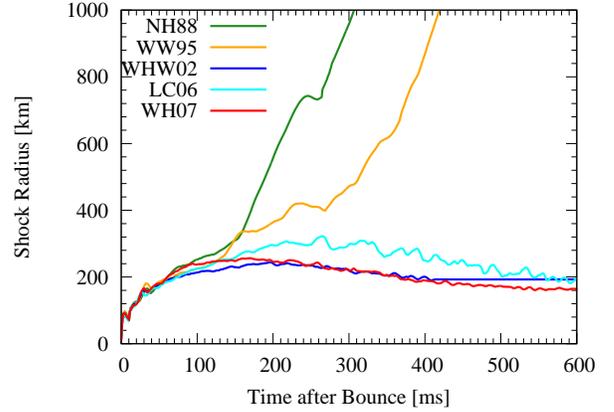}
\caption{Time evolutions of the angle-averaged shock radius for 15
  $M_\odot$ progenitors. NH88 and WW95 produce explosion owing to
  small densities of the envelopes.}
\label{fig:shock_15}
\end{figure}

The shock evolutions for 2D simulations are given in Figure
\ref{fig:shock_15}. The two progenitors, NH88 and WW95, indeed
succeeded in producing shock revival whereas the others failed. This
is a clear demonstration that not the ZAMS mass but the density
structure of progenitor matters for the dynamics of shock revival.
Again, the successful models have turning points that are located more
to the left than the unsuccessful models as seen in Figure
\ref{fig:mdot-lnu_15}.  This is the same conclusion as in the previous
subsection.

\section{Turning point}
\label{sec:critical_turning}

\begin{figure}[tbp]
\centering
\includegraphics[width=0.4\textwidth]{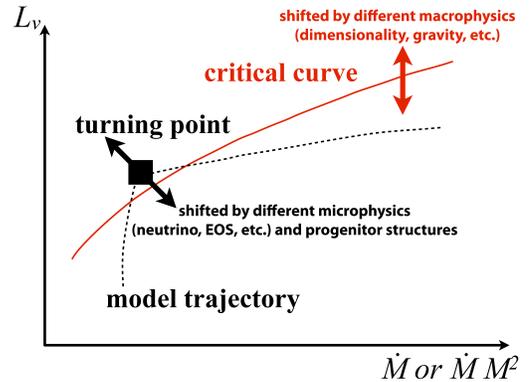} 
\caption{Schematic picture of the critical curve and turning point. If
  the turning point is located above the critical curve and the
  luminosity and mass accretion rate stay in the vicinity of the
  tuning point for a long time, such a model will produce
  explosion. The critical curve is expected to be shifted by
  macrophysics such as dimensionality and the turning point may be
  shifted by microphysics as well as the progenitor structure. The
  critical curve and turning point are also useful to asses the
  influence of a particular physics incorporated.}
\label{fig:schematic}
\end{figure}

In this section, we propose a novel idea to diagnose a possibility of
shock revival using the trajectory in the $\dot M$-$L_{\nu}$ or $\dot M M^2$-$L_{\nu}$ plane
(see Figure \ref{fig:schematic}). This plane is often used to discuss
the critical curve, which divides this plane into two regions: the
region below this line, in which there are steady accretion flows and
the other region above the curve, in which there is no such flow
\citep{burr93}. The latter is therefore interpreted as the region,
where shock revival occurs. The question arises where on the actual
trajectory the critical line is crossed from below?

In Figure \ref{fig:schematic}, we present the typical situation we
found in the majority of our models in the previous sections as a
schematic picture of the trajectory and the critical curve in the
$\dot M$-$L_\nu$ plane.  The red solid line represents the critical
curve and the black dotted line gives a typical trajectory.  As
mentioned already in the preceding sections, there is a point on the
trajectory, at which the slope of the trajectory steepens suddenly as
a consequence of the rapid change in the mass accretion rate there.
This point is referred to as the {\it turning point} in this paper.
It is worth noting that the trajectory is shallower than the critical
curve before the turning is reached and the order is changed
thereafter. Consequently it is obvious that the trajectory can cross
the critical curve if and only if the turning point is located above
the critical curve. It should also be clear that shock revival will be
fizzled if the system evolves rapidly after the turning point, rolling
down the second half of the trajectory and quickly passing the
critical point again. Hence, it is important that the system stays for
a long time around the turning point.

Since the critical curve is a convex and the monotonically increasing
function of the mass accretion rate, the more to the upper left the
turning point is located, the more likely shock revival is to obtain.
Although the critical curve has been well studied by several
groups,\footnote{There are a few attempts to derive the critical curve
  analytically \citep{pejc12,kesh12, jank12}. The impact of properties
  of the nuclear equation of state on the critical curve is also
  studied \citep{couc13a} and is found to be minor compared to the
  dimensionality.}  we emphasize here the importance of the trajectory
as well. In principle, multi-dimensional neutrino-radiation
hydrodynamic simulations, or {\it ab initio} computations, with
detailed neutrino physics and radiative transfer being incorporated
are required to obtain reliable model trajectories.  It has been
demonstrated, however, that one observed effect of
multi-dimensionality in supernova dynamics is to lower the critical
curve \citep{murp08,nord10,hank12}, although the trajectory is also
somewhat modified. Hence, it is expected that 1D simulations will be
sufficient to find approximate locations of turning points and to
infer which models are more likely to explode than others.  1D model
trajectory will also be useful to discuss to what extent particular
ingredients included in simulations (e.g., the nuclear equation of
state, neutrino interactions, scheme to solve the neutrino transfer)
affect the location of the turning point.

In the following, based on the results of our simulations presented so
far, we develop a phenomenological model that connects the density
structure of progenitor just prior to the collapse and the model
trajectory in the $\dot M$-$L_{\nu}$ plane.

\section{Phenomenological model}
\label{sec:phenomenology}

In this section, we construct a phenomenological model.
The purpose is twofold: firstly, it is important to understand
qualitatively why and where the turning point appears; secondly, we
aim to expedite the judgment of which progenitors are likely to
produce shock revival. As mentioned in the previous sections, the
location of the turning point, if any, on the trajectory in the $\dot
M$-$L_\nu$ plane may serve as a sufficient condition for shock revival
if it is located more to the upper left corner. Although the
trajectory evaluated in 1D simulations will be sufficient for this
purpose, the procedure may be simplified even further by the
employment of the phenomenological model. It is not necessary for the
phenomenological models to perfectly reproduce the trajectories
obtained numerically. Instead, it is important that the
turning points are placed at approximately correct positions and that
the relative locations of the turning points for different progenitors
are correctly reproduced. The latter point is particularly important,
since the numerical results contain systematic errors one way or
another.

\begin{figure}[tbp]
\centering
\includegraphics[width=0.45\textwidth]{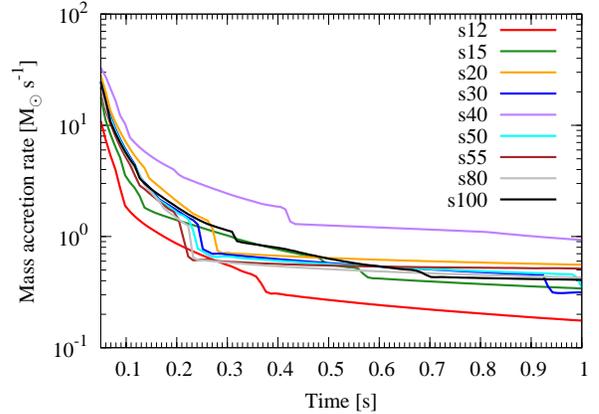}
\caption{Mass accretion rate calculated by the free-falling model.}
\label{fig:mdot}
\end{figure}

In this model the mass accretion rate is evaluated as
\begin{eqnarray}
\dot M&=&\frac{dM}{dt_\mathrm{ff}}\\
&=&\frac{dM}{dr}\left(\frac{dt_\mathrm{ff}}{dr}\right)^{-1},
\label{eq:mdot}
\end{eqnarray}
where $t_\mathrm{ff}$ is the free-fall time, which is defined as a
function of the radius by
\begin{eqnarray}
t_\mathrm{ff}&=&
\alpha \sqrt{\frac{r^{3}}{GM}}\nonumber\\
&\approx& 0.130~\mathrm{s}
\left(\frac{\alpha}{1.5}\right)
\left(\frac{r}{1000~\mathrm{km}}\right)^{3/2}
\left(\frac{M}{M_{\odot}}\right)^{-1/2},
\label{eq:t_ff}
\end{eqnarray}
where $\alpha$ is a parameter introduced to fit to numerical results.
Inverting this relation, we regard the radius as a function of
$t_\mathrm{ff}$. Figure \ref{fig:mdot} shows the mass accretion rates
as a function of $t$, which is identified with $t_\mathrm{ff}$. The
figure should be compared with the bottom panel of Figure \ref{fig:1d}
(note that the vertical scale is different.). It can be seen that the
model reproduces the characteristic features, namely that the mass
accretion rate is high and rapidly decreasing initially and when the
silicon layer accretes onto the PNS completely, it becomes
significantly smaller because of the density drop at the layer
boundary and remains almost constant thereafter (see Appendix
\ref{sec:mdot} for the reason why the mass accretion rate becomes
constant at late times).

Although there have been several approximate functional forms,
e.g. $e^{-t/\tau}$ \citep{jank96}, proposed for the total neutrino
luminosity as a function of time, we employ the following form based
on the diffusion time scale:
\begin{eqnarray}
L_{\nu}(t)=\frac{L_\mathrm{diff}}{1+t/t_\mathrm{diff}},
\label{eq:Lnu1}
\end{eqnarray}
where $L_\mathrm{diff}=E_\mathrm{int}/t_\mathrm{diff}$ is the
diffusion luminosity with
$E_\mathrm{int}=(3/5)GM_\mathrm{PNS}^{2}/R_{\nu}$ being the internal
energy stored inside the PNS and $t_\mathrm{diff}$ being the diffusion
timescale defined shortly, and $M_\mathrm{PNS}$ and $R_{\nu}$ are the
mass and radius of the PNS, respectively. Again identifying
$t_\mathrm{ff}$ with $t$ results in the following expression,
\begin{eqnarray}
L_{\nu}=\frac{E_\mathrm{int}}{t_\mathrm{ff}+t_\mathrm{diff}}.
\label{eq:Lnu}
\end{eqnarray}
The diffusion time $t_\mathrm{diff}$ can be evaluated as (see Appendix
\ref{sec:diffusion} for the derivation)
\begin{eqnarray}
t_\mathrm{diff}&=&
\frac{3\sigma}{4\pi cm_\mathrm{p}}\frac{M}{R_{\nu}}\nonumber\\
&\approx& 0.402~\mathrm{s}
\left(\frac{\varepsilon_{\nu,\mathrm{PNS}}}{57~\mathrm{MeV}}\right)^{2}
\left(\frac{M}{M_{\odot}}\right)
\left(\frac{R_{\nu}}{50~\mathrm{km}}\right)^{-1},
\label{eq:t_diff}
\end{eqnarray}
where $\sigma$ is the cross section of neutrino-nucleon scattering,
which is given as
$\sigma(\varepsilon_{\nu})\approx\sigma_{0}(\varepsilon_{\nu}/m_\mathrm{e}c^{2})^{2}$
with $\sigma_{0}=1.705\times 10^{-44}~\mathrm{cm}^{2}$,\footnote{There
  are coefficients of $O(1)$, which are neglected for simplicity
  \citep[see][for more details]{burr06b}.} the electron mass
$m_\mathrm{e}$, and the neutrino energy $\varepsilon_{\nu}$; the
proton mass is denoted by $m_\mathrm{p}$. Moreover,
$\varepsilon_{\nu,\mathrm{PNS}}$ is a characteristic energy of
neutrinos inside the PNS.\footnote{The characteristic value employed
  here seems rather large compared with the commonly used value $\sim
  10$ MeV. This is because the former represents the average energy
  inside the PNS, where the matter temperature is a few tens MeV and,
  as a consequence, the neutrino average energy becomes several tens
  MeV. On the other hand, the latter value reflects the matter
  temperature at the neutrinosphere, $O(1)$ MeV.} The mass of PNS
increases as matter accretes and can be expressed as a function of
time by the use of the free fall time.

The idea underlying Eq. (\ref{eq:Lnu}) is that the material initially
located at $r$ falls onto the PNS in its free-fall time and the
gravitational energy is converted to the internal energy, which is
finally radiated as neutrinos in the diffusion time.  We can then
evaluate the neutrino luminosity as Eq. (\ref{eq:Lnu}).
This phenomenological model is consistent with the finding by
\citet{fisc09} and \citet{muel14} that the neutrino luminosity seems
to be regulated by the smaller of the accretion and diffusion
luminosities. Indeed, since $L_\mathrm{acc}\sim
E_\mathrm{int}/t_\mathrm{ff}$ and $L_\mathrm{diff}\sim
E_\mathrm{int}/t_\mathrm{diff}$, $L_\mathrm{acc}<L_\mathrm{diff}$ for
$t_\mathrm{ff}>t_\mathrm{diff}$, and vise versa.

\begin{figure}[tbp]
\centering
\includegraphics[width=0.45\textwidth]{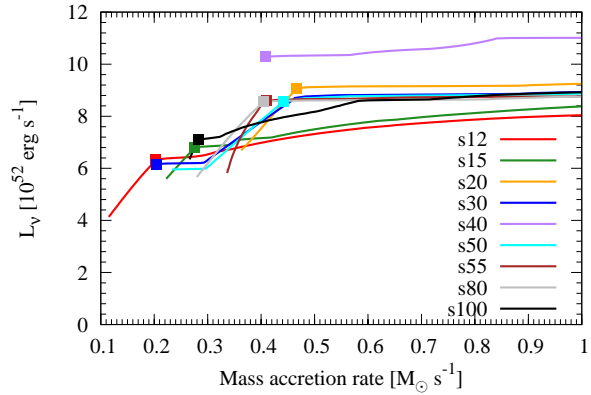}
\caption{Phenomenological model for neutrino luminosity as a function
  of mass accretion rate.}
\label{fig:mdot-lnu-ana}
\end{figure}

Figure \ref{fig:mdot-lnu-ana} presents the model trajectories in the
$\dot M$-$L_{\nu}$ plane obtained this way.  In this plot, we employ
$\alpha=1.5$, $\varepsilon_{\nu,\mathrm{PNS}}=57$ MeV, and
$R_{\nu}=50$ km in Equations (\ref{eq:t_ff}) and (\ref{eq:t_diff}).
The turning points, where the slope of trajectory changes rapidly, are
clearly produced in most cases.  The comparison between the
phenomenological model and the numerical results is given in Appendix
\ref{sec:comparison} (see Fig. \ref{fig:comparison}).  Filled squares
shown in Figure \ref{fig:mdot-lnu-ana} represent the points on each
trajectory, which will be the most favorable for shock revival and are
determined so that the value of the ratio of the calculated $L_\nu$ to
the critical luminosity given by \cite{burr93} as
\begin{equation}
L_\mathrm{BG93}^\mathrm{crit}=5\times 10^{52} \mathrm{erg~s^{-1}}
\left(\frac{\dot M}{1.1~M_\odot~\mathrm{s}^{-1}}\right)^{1/2.3}
\label{eq:bg}
\end{equation}
should be maximum. Note that, strictly speaking, Eq. (\ref{eq:bg}) is
valid only for the luminosity of electron-type neutrino, $L_{\nu_e}$,
with the temperature of $kT_{\nu_e}=4.5$ MeV. Here the Boltzmann
constant is denoted by $k$. We believe, however, that other
expressions will not change the following discussions.  It can be
found that the filled squares coincide with the turning points,
whenever they exist.  For model s55, for example, this point occurs at
$\dot M\approx 0.4M_\odot$ s$^{-1}$, which is consistent with the
numerical result (see Figure \ref{fig:mdot-lnu}), and $L_\nu$ is a
rapidly increasing function of $\dot M$ for smaller mass accretion
rates whereas it increases rather slowly for larger mass accretion
rates. As mentioned before this drastic change comes from the
transition of accreting layers, i.e., from silicon to oxygen layers.
A comparison between critical curve of \cite{burr93} and turning
points presented here is given in Appendix \ref{sec:bg93}, which shows
that the critical curve is basically compatible with turning point
locations and numerical results, i.e. success and failure of
explosion, but with some exemptions. Next, we try to improve our
critical curve in a similar plot.

\begin{figure}[tbp]
\centering
\subfigure[Ratio between model trajectories and critical curve (Eq. \ref{eq:Lcrit_j})]{\includegraphics[width=0.45\textwidth]{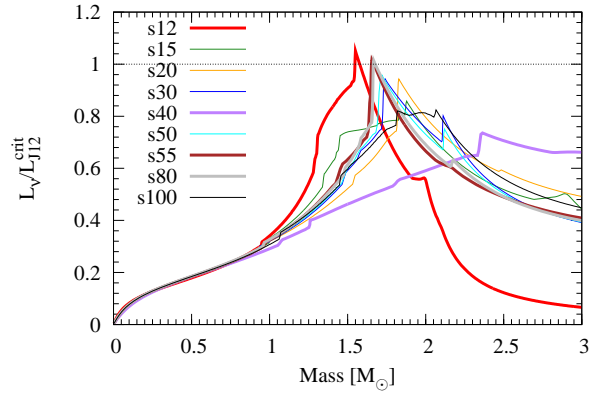}\label{fig:m-ratio}}
\subfigure[The locations of the turning points in $\dot MM^2$-$L_\nu$ plane.]{\includegraphics[width=0.45\textwidth]{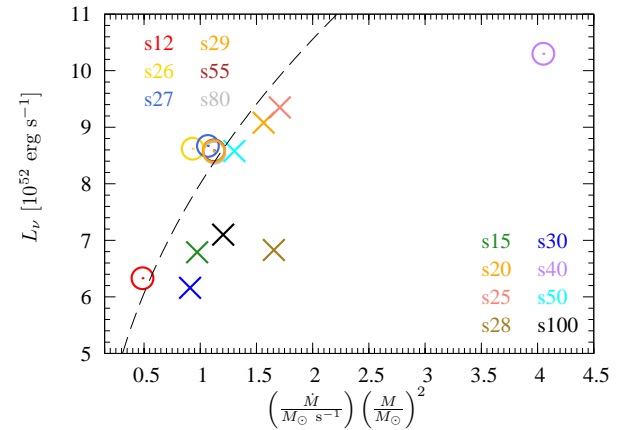}\label{fig:mdm2-lnu}}
\caption{Ratio between model trajectories and critical curve
  (Eq. \ref{eq:Lcrit_j}) as a function of mass coordinate for each
  progenitor models (a) and the locations of the turning points in
  $\dot MM^2$-$L_\nu$ plane (b). These panels are based on
  phenomenological model. In panel (b), additional models (s25, s26,
  s27, s28, and s29) presented in Appendix \ref{sec:others} are also
  plotted for references. As Figure \ref{fig:mdot-lnu-explode},
  exploding models are shown as open circles and non-exploding models
  are shown as crosses.}
\end{figure}

Guided
by a simple analytical derivation of the critical curve from
\cite{jank12} (Section 4.3.4 in this review), the critical curve
depends not only on $\dot M$, but also NS mass $M_\mathrm{NS}$ as
$L^\mathrm{crit}_\mathrm{J12}\propto \dot
M^{2/5}M_\mathrm{NS}^{4/5}$.\footnote{One can find a more detailed
  expression of the critical curve in \cite{pejc12,muel15a}.} Note
that in this formula $M_\mathrm{NS}$ denotes the point mass
determining the gravitational potential so that it should be replaced
with enclosed mass for each mass element. In Figure \ref{fig:m-ratio},
we show the distribution of the ratio between total neutrino
luminosity calculated by the phenomenological model (Eq. \ref{eq:Lnu})
and the critical curve by \cite{jank12} as
\begin{equation}
L^\mathrm{crit}_\mathrm{J12}=8\times 10^{52}\mathrm{erg~s^{-1}}
\left(\frac{\dot M}{M_\odot}\right)^{2/5}
\left(\frac{M}{M_\odot}\right)^{4/5}.
\label{eq:Lcrit_j}
\end{equation}
The overall factor is chosen to divide exploding and non-exploding
models as follows. 
This quantity, $L_\nu/L_\mathrm{crit}$, represents how long model
trajectories are distant from critical curve, which is useful to
predict model exploitability.
In this figure, exploding models (s12, s40, s55, and s80) are
indicated with thick lines and failing models (s15, s20, s30, s50, and
s100) are indicated with thin lines, respectively. One can find three
exploding models (s12, s55, and s80) exceed unity and other models
stay below unity for the whole regime.  The exceptional behavior of
s40 originates from omission of neutrino energy dependence. As
indicated by more realistic expressions of the critical curve
\citep[e.g.,][]{pejc12,muel15a}, the higher RMS energy
$\sqrt{\bracket{\epsilon_\nu^2}}$ leads to smaller critical
luminosity. The RMS energy of s40 is typically higher than in other
models, namely, by $\sim 10$\% so that its critical luminosity is
smaller, by $\sim 20$\%. Since the peak value of
$L_\nu/L^\mathrm{crit}_\mathrm{J12}$ of s40 is about $\sim 0.8$, the
inclusion of RMS energy really improves the situation. Note that the
RMS energy of the other models are very similar (see Figure
\ref{fig:neutrino}) so that the critical curve of other models does
not change considerably.  The modeling of RMS energy is currently not
feasible.  However, most models (except a model with a significantly
low or high mass accretion rate) imply similar RMS energies \citep[see
  also][]{lieb03} and their dependence on the progenitor structure is
lower than luminosities. Thus, the critical curve given by
Eq. (\ref{eq:Lcrit_j}) works quite well to predict the
explosion/failure from the progenitor model alone.
Note also that these peaks in Figure \ref{fig:m-ratio} coincide with
the position of turning points, even though we include the term of
enclosed mass in the expression of the critical curve
(Eq. \ref{eq:Lcrit_j}).

As described in Section \ref{sec:critical_turning}, a comparison
between the turning points and the critical curve gives a useful
criterion to judge which progenitors are likely to produce shock
revival, which is shown in Figure \ref{fig:mdm2-lnu}. The horizontal
axis is replaced from $\dot{M}$ to $\dot{M}M^2$ following
\cite{jank12}. The exploding models are denoted by open circles and
the failed models are denoted by crosses.  The comparison works quite
well for most of the models with only an exception of s40, as
discussed above. A similar plot is shown in Figure
\ref{fig:mdot-lnu-explode} in Appendix \ref{sec:bg93}, which has
different horizontal quantity, $\dot{M}$.  Note that Figure
\ref{fig:mdm2-lnu} includes additional circles and crosses of s25,
s26, s27, s28, and s29, which are presented in Appendix
\ref{sec:others}. It can be seen that this panel gives more compatible
distributions of turning points and position of critical curve with 2D
simulations (i.e. exploding or non-exploding) than Figure
\ref{fig:mdot-lnu-explode}.

It is true that the trajectories and the critical curve we obtained in
this paper will change when the numerics are improved one way or
another, but it should be stressed that the following methodology
should be applied to any combination of a numerical code and to a
collection of progenitors: first we perform a small number of 1D
simulations to fix the parameters in the phenomenology and then apply
the latter to a large number of progenitors to predict which ones are
more likely to explode than others. As an example, we presented the
locations of turning points obtained that way for other progenitor
models from \cite{woos07} in Appendix \ref{sec:others}.

\section{Summary and discussions}
\label{sec:summary}

In this paper, we performed neutrino-radiation hydrodynamic
simulations in spherical symmetry (1D) and in axial symmetry (2D) for
different progenitor models by \cite{woos07} from 12 $M_{\odot}$ to
100 $M_{\odot}$. We found that all 1D runs failed to produce an
explosion and several 2D runs succeeded. The difference in the shock
evolutions can be mainly ascribed to different mass accretion
histories, which are determined by the density structures of
progenitors.
For the majority of the models we studied in this paper we found that
the mass accretion rate changes its nature suddenly at a certain
point: in the earlier phase it is very high and decreasing quickly
whereas it settles to a nearly constant rate in the later phase. In
the $\dot M$-$L_\nu$ plane this transition point, which we called the
turning point, marks the point, where the slope of the trajectory
changes rapidly: the trajectory is shallower than the critical curve
in the earlier phase and becomes steeper than that thereafter. This
means that the trajectory crosses the critical curve from below around
the turning point if it ever occurs and that shock revival is most
likely to take place there. It is hence obvious that we can employ the
position of the turning point as a diagnostic for shock revival. It
should be noted, however, that the turning point does not always seem
to exist and that the above criterion should be regarded as a
sufficient condition.  In addition, we developed a phenomenological
model to approximately estimate trajectories in the $\dot M$-$L_\nu$
plane. This model utilizes the initial density structure of progenitor
alone and reproduces the locations of turning points reasonably
well. Based on these results, we suggest the following usage of the
phenomenological model: perform a small number of 1D simulations to
fix the free parameters in the model and apply the result to a large
number of progenitors to infer which progenitors are more likely to
explode.  It should be noted that the main effects of the intrinsic
multi-dimensionality of supernova dynamics is mainly to shift the
critical curve parallelly downward and its influences on the
trajectory are limited.

The phenomenological model depends on the underlying 1D simulations,
and hence on the numerics and input microphysics adopted therein. The
important thing, however, is that the methodology is applicable to any
combination of a numerical code and a collection of progenitor models.
Some comments on the simulations by other groups follow.
\cite{brue13,brue14} reported successful explosions for 12, 15, 20 and
25 $M_\odot$ progenitor models of \cite{woos07}, which are identical
to those employed in this paper.
\cite{dole15} adopted the same progenitors in their simulations and
found no explosion. Both of them employed the flux-limited diffusion
approximation for neutrino transfer. \cite{dole15} suggested that the
reason for this discrepancy might be the use of the ray-by-ray
approximation by \cite{brue13}.
We obtained an explosion for the 12 $M_\odot$ model, but not for the
15 and 20 $M_\odot$ models, on the other hand, which indicates that
our simulations fall somewhere between them.  Note that we also
employed the ray-by-ray approximation.

\begin{figure}
\centering
\includegraphics[width=0.45\textwidth]{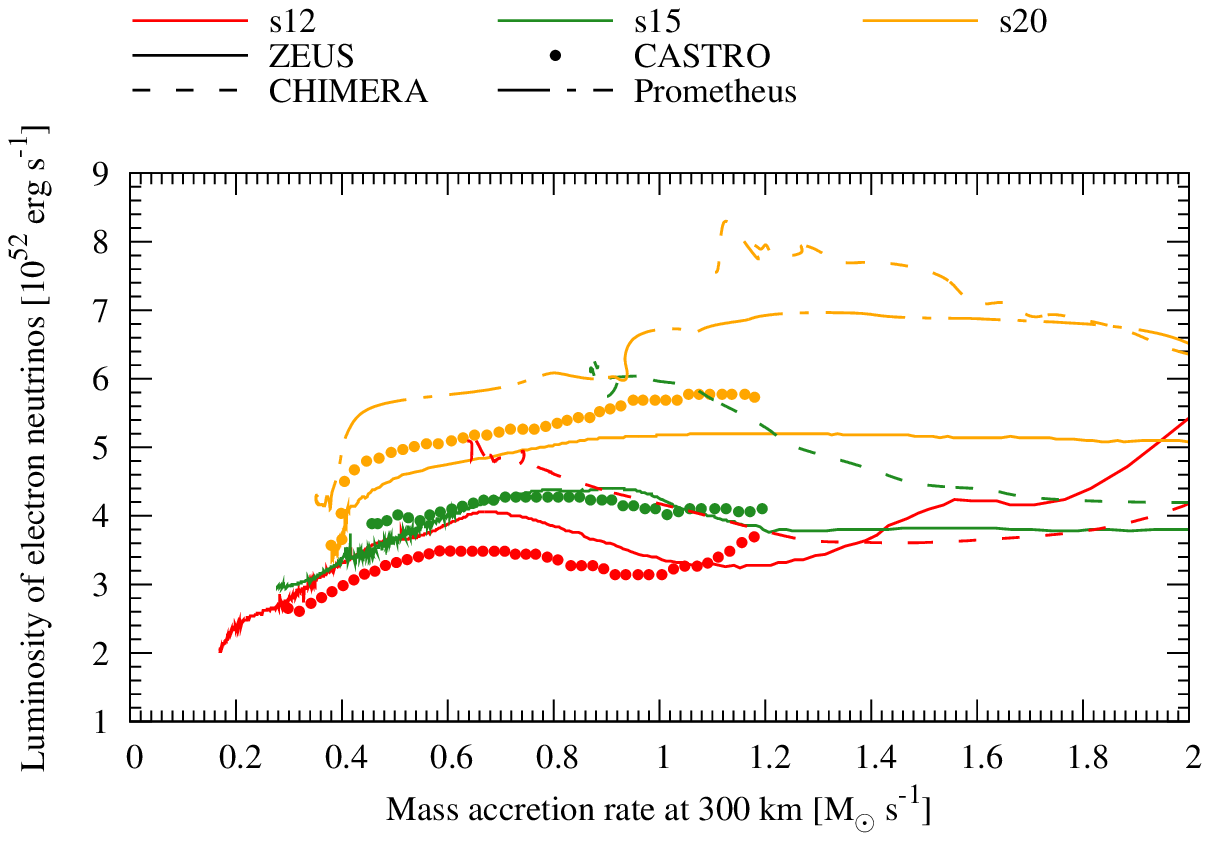}
\caption{Comparison of model trajectories of this study (solid lines),
  \cite{dole15} (dots; taken from Figure 7 of their paper),
  \cite{brue14} (dashed lines; made from Figure 1 of their paper), and
  \cite{mels15b} (dot-dashed lines, only for s20; made from Figures 3
  and 4 of their paper). Note that trajectories of \cite{brue14} are
  plotted up to 150 ms after the bounce, while those of this study,
  \cite{dole15}, and \cite{mels15b} are plotted up to several hundreds
  of milliseconds postbounce.}
\label{fig:mdot-lnu_compare}
\end{figure}

Figure \ref{fig:mdot-lnu_compare} compares the trajectories in the
$\dot M$--$L_{\nu_e}$ plane for these models.\footnote{Note that not
  in $\dot M$--$L_{\nu}$ plane.} The solid lines are the trajectories
for the models in this paper, the dots are obtained from Figure 7 of
\cite{dole15}, the dashed lines are drawn based on Figure 1 of
\cite{brue14}, and the dot-dashed lines (only for s20) is generated
from Figures 3 and 4 of \cite{mels15b}. Our trajectories are closer to
those of \cite{dole15} and \cite{mels15b}, while the trajectories of
\cite{brue14} are apparently running in the higher luminosity regime,
which may have lead to the successful explosions in their simulations.
Note that \cite{mels15b} also obtained an explosion for s20, which
might also be a consequence of higher neutrino luminosity at the
turning point. However, the position of the critical curve depends on
physical ingredients implemented in each code.

Recently, \cite{naka14} performed 2D simulations for a large number of
progenitor models of 10.8-75.0 $M_\odot$ taken from \cite{woos02},
which are incidentally different from our progenitor models, and
obtained explosions for all the models. Although their code, employing
the IDSA and ray-by-ray approximation for neutrino transfer, is quite
similar to ours, it is not the same.  This is because their
hydrodynamics module employs a different shock-capturing scheme and
that may be the source of differences in the outcomes.

Finally, we comment on the assumptions adopted in this study. Firstly,
we performed 2D simulations although it is well known that axial
symmetry leads to some hydrodynamic features that are qualitatively
different from those in three dimensions (3D)
\citep{couc13b,hank13,hand14,taki14}. The critical curve in 3D is
observed to shift from that in 2D \citep{nord10,hank12} although the
magnitude and its dependence on the spacial resolution is still
controversial.  Incidentally, 3D hydrodynamic simulations with
spectral neutrino transfer are currently still computationally too
expensive to perform as a systematic study like the one in this
paper. Secondly, the microphysics used in this study is not so
elaborate as other numerical studies \citep{muel12b,brue13} and our
critical curve may be different from theirs. The phenomenological
model we proposed in this paper should hence be applied to more
sophisticated numerical simulations, which will be carried out as a
future project.

\acknowledgements 

We thank A. Heger, M. Limongi, and K. Nomoto for providing progenitor
models, M. Tanaka for fruitful discussions, and S. Veith for
proofreading. The numerical computations in this study were partly
carried out on XT4 and XC30 at CfCA in NAOJ and SR16000 at YITP in
Kyoto University. This study was supported in part by the Grant-in-Aid
for Scientific Research (Nos. 25103511, 26870823, 23540323, 23340069,
24103006, 26707013, and 24244036), JSPS postdoctoral fellowships for
research abroad, MEXT SPIRE, and JICFuS.

\appendix

\section{Conditions for constant $\dot M$}
\label{sec:mdot}

In this section, we give a simple explanation of why we obtain almost
constant mass accretion rates at late times for most of the
progenitors in this study. We assume the following density structure,
\begin{equation}
\rho(r)=\rho_0\left(\frac{r}{R}\right)^{-n},
\end{equation}
where $R$ is a core radius and $\rho_0$ is the density at $r=R$. The
mass coordinate is given by
\begin{equation}
M(r)=M_0+\int^r_R 4\pi r'^2\rho(r') dr',
\end{equation}
where $M_0$ is the mass coordinate at $r=R$. Then, the mass accretion
rate is estimated as
\begin{equation}
\dot M
=\frac{dM}{dt_\mathrm{ff}}
=\frac{dM}{dr}\left(\frac{dt_\mathrm{ff}}{dr}\right)^{-1},
\end{equation}
where $t_\mathrm{ff}=\sqrt{r^3/GM(r)}$ is the free fall timescale.
Short calculations give
\begin{equation}
\frac{dM}{dr}=4\pi r^2\rho_0\left(\frac{r}{R}\right)^{-n},
\end{equation}
and
\begin{equation}
\frac{dt_\mathrm{ff}}{dr}
=\frac{1}{2}\sqrt{\frac{r}{GM(r)}}\left\{3-4\pi r^2\rho_0\left(\frac{r}{R}\right)^{-n}\frac{r}{M(r)}\right\}^{-1}.
\end{equation}
Suppose that $M(r)\approx M_0$, i.e., the central accelerator's mass
is dominant and $r\gg R$. Then, $dt_\mathrm{ff}/dr\approx
(3/2)\sqrt{r/GM(r)}$ and we obtain
\begin{equation}
\dot M\approx \frac{8\pi\rho_0 R^n\sqrt{GM_0}}{3} r^{\frac{3}{2}-n},
\end{equation}
which becomes constant if $n=3/2$.
If the mass at $r>R$, that is the mass of accreting matter, is
dominant, we get
\begin{equation}
M(r)\approx \frac{4\pi\rho_0 R^n}{3-n}r^{3-n},
\end{equation}
which leads to
\begin{equation}
\frac{dt_\mathrm{ff}}{dr}\approx\frac{n}{2}\sqrt{\frac{3-n}{4\pi G\rho_0 R^n}}r^{\frac{n}{2}-1}.
\end{equation}
Then we obtain
\begin{equation}
\dot M\approx \frac{2}{n}\sqrt{\frac{(4\pi\rho_0 R)^3G}{3-n}} r^{3-\frac{3}{2}n},
\end{equation}
which is again constant for $n=2$. The progenitor models used in this
study realize $n\approx 2$ for the oxygen layer so that the latter
case is valid for them.

\section{Diffusion timescale}
\label{sec:diffusion}

Here, we obtain a useful expression of the diffusion timescale for
neutrinos in a uniform density sphere of radius $R_\nu$, which is
meant to be a rough approximation to a PNS. The diffusion timescale is
given by
\begin{equation}
t_\mathrm{diff}=\frac{\tau_\nu R_\nu}{c},
\label{eq:tdiff}
\end{equation}
where $\tau_\nu$ is the optical depth of the sphere, which is 
\begin{eqnarray}
\tau_\nu=&\int^{R_\nu}_0 dr \displaystyle\frac{\rho \sigma}{m_\mathrm{p}}\\
=&\displaystyle\frac{3\sigma}{4\pi m_\mathrm{p}}\frac{M}{R_\nu^2}.
\label{eq:tau}
\end{eqnarray}
Here we used $M=4\pi \rho R_\nu^3/3$.  By combining
Eqs. (\ref{eq:tdiff}) and (\ref{eq:tau}), we get
\begin{equation}
t_\mathrm{diff}=\frac{3\sigma}{4\pi c m_\mathrm{p}}\frac{M}{R_\nu}.
\end{equation}

\section{Comparison between phenomenological and numerical models}
\label{sec:comparison}

\begin{figure}[tbp]
\centering
\includegraphics[width=0.45\textwidth]{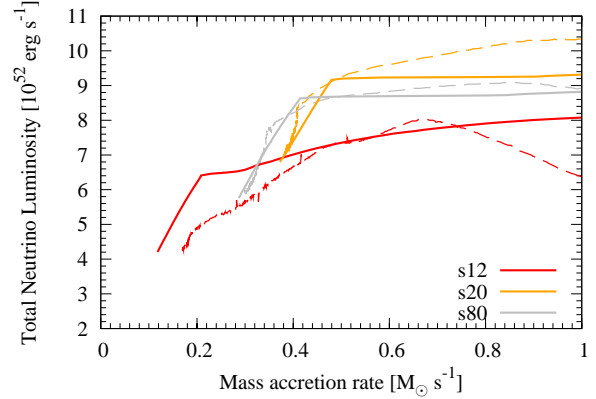}
\caption{Comparison between numerical simulation (dashed lines) and
  the phenomenological model (solid lines) for selected models.}
\label{fig:comparison}
\end{figure}

In this section, we show the comparison between the phenomenological
model introduced in Section \ref{sec:phenomenology} and the numerical
results presented in Section \ref{sec:1D}. Figure \ref{fig:comparison}
presents the model trajectories for models s12, s20, and s80. Solid
curves show the model trajectories obtained with the phenomenological
models and dashed curves display the trajectories given by the
numerical simulations. It can be found that for s80 these lines agree
very well, while for s12 the phenomenological model fails to reproduce
the numerical result. As for s20, there is a discrepancy between two
lines at high mass accretion rates, whereas the turning point is
almost perfectly reproduced. For models s30, s50, and s55, which have
clear turning points in the simulations they are reproduced reasonably
well by the phenomenological model. However, for models s15, s40, and
s100, we cannot fit well. These results indicate that the
phenomenological model is useful for progenitors that have clear
turning points, i.e., progenitors with a large density jump between
the silicon and oxygen layers.

\section{Turning Points in $\dot M$-$L_\nu$ Plane and Critical Curve}
\label{sec:bg93}

\begin{figure}[tbp]
\centering
\includegraphics[width=0.45\textwidth]{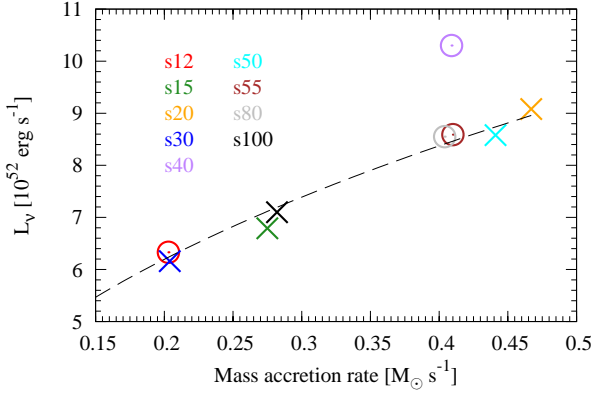}
\caption{ The locations of the turning points of the investigated
  models are represented as open circles for exploding models and
  crosses for non-exploding models in 2D simulations given in Section
  \ref{sec:2D}. The dashed line is $L_\nu=13\times 10^{52}
  \mathrm{erg~s^{-1}}({\dot
    M}/{1.1~M_\odot~\mathrm{s}^{-1}})^{1/2.3}$, which seems to divide
  the phase space into exploding and non-exploding regions.  }
\label{fig:mdot-lnu-explode}
\end{figure}

In Figure \ref{fig:mdot-lnu-explode} the locations of the turning
points obtained by the phenomenological model are plotted for
different progenitors as open circles for exploding models and as
crosses for non-exploding models in 2D simulations described in
Section \ref{sec:2D}. The dashed line is given by $L_\nu=13\times
10^{52} \mathrm{erg~s^{-1}}({\dot
  M}/{1.1~M_\odot~\mathrm{s}^{-1}})^{1/2.3}$, which we fit to roughly
divide the exploding models from the non-exploding ones. These
critical luminosities are slightly higher than those obtained by
\citet{murp08} with the light-bulb approximation:
$L_\nu\approx10\times 10^{52} \mathrm{erg~s^{-1}}({\dot
  M}/{1.1~M_\odot~\mathrm{s}^{-1}})^{1/2.3}$. The discrepancy should
be ascribed to the difference in the numerical treatments of neutrino
transfer. It should be noted, however, that this fit is not perfect
and, for instance, s20 lies above the line. We stress that the
location of the turning point is indeed a useful measure when one
attempts to infer the possibility of explosion from the progenitor
structure alone.

\section{Other Progenitors}
\label{sec:others}

\begin{figure}[tbp]
\centering
\includegraphics[width=0.45\textwidth]{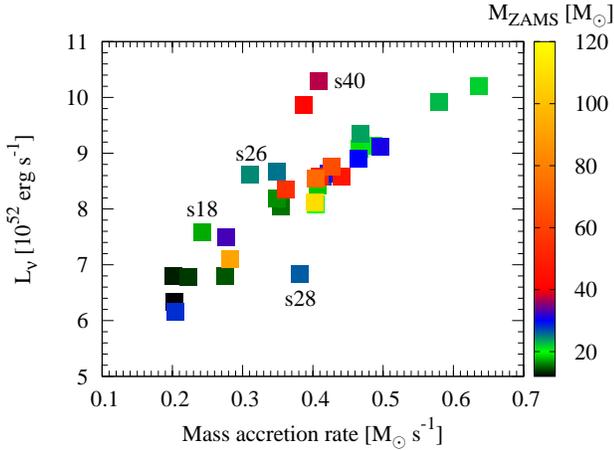} 
\caption{The location of turning points for all progenitor models of
  \cite{woos07}, with color bar denoting the ZAMS mass. The typical
  progenitor models are labeled.}
\label{fig:mdot-lnu-models}
\end{figure}

We show the turning points for all 32 progenitors from \cite{woos07}
in Figure \ref{fig:mdot-lnu-models}. The turning point is defined for
each model to be the point in the $\dot M$-$L_\nu$ plane, at which the
ratio of $L_\nu/L_\mathrm{BG93}^{crit}$ takes the maximum value on the
trajectory. The values of max($L_\nu/L_\mathrm{BG93}^\mathrm{crit}$)
at the turning points are summarized in Table \ref{tab:ratio}, in
which the compactness parameters $\xi_{1.5}$, $\xi_{1.75}$, and
$\xi_{2.5}$ at the precollapse phase are also given (see
\citealt{sukh14} for the relations of these quantities with the
compactness parameters defined at the bounce). Note that these
parameters can be also evaluated only from the density structure of
progenitor and no simulation is required.

There is no clear correlation between the maximum values of
($L_\nu/L_\mathrm{BG93}^\mathrm{crit}$) and the compactness
parameters. The critical value of
max($L_\nu/L_\mathrm{BG93}^\mathrm{crit}$) that divides exploding from
non-exploding models may be set at is $\sim$ 2.18 because models s55
and s80 (max($L_\nu/L_\mathrm{BG93}^\mathrm{crit}$)=2.19) explode,
while s20 (max($L_\nu/L_\mathrm{BG93}^\mathrm{crit}$)=2.18) fails. It
is hence true that max($L_\nu/L_\mathrm{BG93}^\mathrm{crit}$) is very
useful in judging whether a particular model is likely to explode
before doing detailed simulations. In addition to
max($L_\nu/L_\mathrm{BG93}^\mathrm{crit}$), we show
max($L_\nu/L_\mathrm{J12}^\mathrm{crit}$) as well in this table. The
explosion criterion of this indicator is unity. Therefore, 12 models
of all 32 models are expected to make explosions with our current
numerical code.

\begin{figure}[tbp]
\centering
\includegraphics[width=0.45\textwidth]{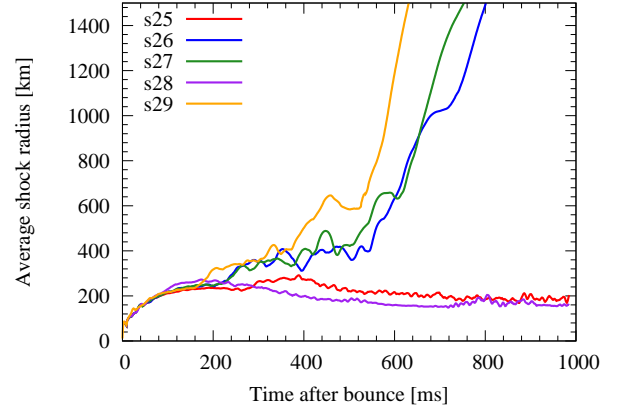} 
\caption{Shock trajectories of models s25, s26, s27, s28, and
  s29. Average shock radius is measured in postbounce time.}
\label{fig:shock_s25-29}
\end{figure}

To confirm robustness of the criterion, we perform additional
simulations of different progenitor models. From Table \ref{tab:ratio}
we pick up from s25 to s29, which include representative models giving
maximum and minimum values for
max($L_\nu/L_\mathrm{BG93}^\mathrm{crit}$), i.e. s26 and s28. The
resultant shock evolution of 2D simulations is shown in Figure
\ref{fig:shock_s25-29}, which displays that the explosion of three
models (s26, s27, and s29) and the failing of two others (s25 and
s28).
Although the first indicator,
max($L_\nu/L_\mathrm{BG93}^\mathrm{crit}$), does not perfectly fit
this result since there is inversion hierarchy between s25 (2.25 and
non-exploding) and s29 (2.17 and exploding), the second indicator,
max($L_\nu/L_\mathrm{J12}^\mathrm{crit}$), solves this contradictory,
i.e. s25 is below unity (0.938) and s29 is above unity (1.018).

\begin{figure}[tbp]
\centering
\includegraphics[width=0.45\textwidth]{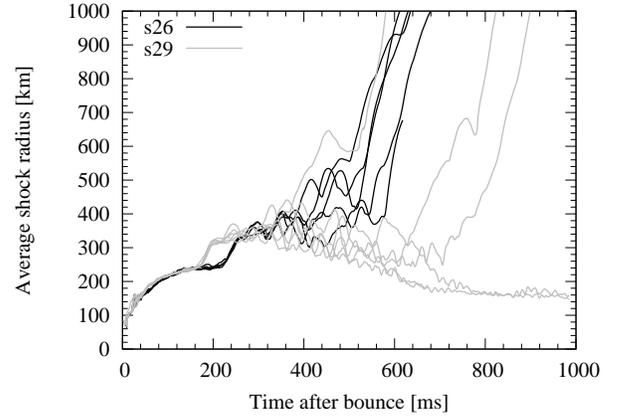} 
\caption{Dependence of shock trajectories on initial seed
  perturbation. Black lines present s26 and gray lines indicate s29.}
\label{fig:shock_s26_s29}
\end{figure}

The value of the second indicator for s29 implies that this model is
rather marginal. To check this marginality, we performed additional
simulations with different initial perturbations. In addition, we
performed additional simulations for s26 as well, because it is
expected to explode more robustly, as indicated by
max($L_\nu/L_\mathrm{BG93}^\mathrm{crit}$) and
max($L_\nu/L_\mathrm{J12}^\mathrm{crit}$). Figure
\ref{fig:shock_s26_s29} presents shock evolutions with different seed
perturbations (0.1\% in density). One can see that all s26 simulations
explode at almost the same time, while half of s29 simulations explode
at very different times.  This result implies that models with high
values of indicators explode robustly and those with marginal values
sometimes fail or explode at different times \citep[see][for the
  explosion fraction]{hori14}. Note that the diversity of explosion
time is narrowed in 3D simulations \citep{taki14,hand14}.

\begin{table*}[tbp]
\centering
\caption{Properties of all progenitors}
\label{tab:ratio}
\begin{tabular}{ccccccc}
\hline
Model & $M_\mathrm{ZAMS}$ & max($L_\nu/L_\mathrm{BG93}^\mathrm{crit}$) & max($L_\nu/L_\mathrm{J12}^\mathrm{crit}$) & $\xi_{1.5}^\mathrm{pre}$ & $\xi_{1.75}^\mathrm{pre}$ & $\xi_{2.5}^\mathrm{pre}$ \\
& ($M_\odot$) &  &  &   & \\
\hline
s12 & 12 &     2.19 &    1.051 &    0.617 &    0.235 &    0.023 \\
s13 & 13 &     2.36 &    1.064 &    0.869 &    0.370 &    0.067 \\
s14 & 14 &     2.25 &    0.972 &    0.857 &    0.502 &    0.128 \\
s15 & 15 &     2.06 &    0.858 &    0.882 &    0.549 &    0.181 \\
s16 & 16 &     2.19 &    1.095 &    0.792 &    0.333 &    0.150 \\
s17 & 17 &     2.24 &    1.093 &    0.877 &    0.374 &    0.168 \\
s18 & 18 &     2.43 &    0.996 &    0.961 &    0.656 &    0.194 \\
s19 & 19 &     2.15 &    1.006 &    0.962 &    0.517 &    0.177 \\
s20 & 20 &     2.18 &    0.943 &    1.003 &    0.771 &    0.286 \\
s21 & 21 &     2.07 &    1.060 &    0.696 &    0.323 &    0.143 \\
s22 & 22 &     2.16 &    0.929 &    1.001 &    0.783 &    0.289 \\
s23 & 23 &     2.14 &    0.828 &    0.998 &    0.870 &    0.434 \\
s24 & 24 &     2.17 &    0.859 &    1.013 &    0.859 &    0.398 \\
s25 & 25 &     2.25 &    0.938 &    1.008 &    0.821 &    0.331 \\
s26 & 26 &     2.48 &    1.101 &    0.968 &    0.641 &    0.234 \\
s27 & 27 &     2.37 &    1.048 &    0.993 &    0.677 &    0.257 \\
s28 & 28 &     1.79 &    0.716 &    0.987 &    0.596 &    0.272 \\
s29 & 29 &     2.17 &    1.018 &    0.965 &    0.534 &    0.225 \\
s30 & 30 &     2.13 &    0.944 &    1.006 &    0.688 &    0.218 \\
s31 & 31 &     2.17 &    0.983 &    0.995 &    0.617 &    0.219 \\
s32 & 32 &     2.15 &    0.942 &    0.999 &    0.750 &    0.253 \\
s33 & 33 &     2.14 &    0.915 &    1.003 &    0.783 &    0.284 \\
s35 & 35 &     2.26 &    0.861 &    1.010 &    0.846 &    0.360 \\
s40 & 40 &     2.62 &    0.736 &    0.980 &    0.876 &    0.547 \\
s45 & 45 &     2.58 &    0.743 &    0.982 &    0.875 &    0.516 \\
s50 & 50 &     2.12 &    0.958 &    0.999 &    0.643 &    0.222 \\
s55 & 55 &     2.19 &    1.018 &    0.980 &    0.564 &    0.239 \\
s60 & 60 &     2.24 &    1.066 &    0.939 &    0.451 &    0.175 \\
s70 & 70 &     2.20 &    0.987 &    0.989 &    0.663 &    0.233 \\
s80 & 80 &     2.19 &    1.014 &    0.965 &    0.550 &    0.210 \\
s100 & 100 &     2.13 &    0.824 &    0.989 &    0.702 &    0.245 \\
s120 & 120 &     2.09 &    0.842 &    0.911 &    0.454 &    0.171 \\
\hline
\end{tabular}
\end{table*}

\end{document}